\documentclass{aa}  

\usepackage{graphicx}
\usepackage{txfonts}

\usepackage{graphicx,rotating,amssymb,amsmath}
\usepackage{booktabs}
\usepackage{subfig}
\usepackage{float,capt-of}
\usepackage{natbib}
\usepackage{multirow}
\usepackage[english]{babel}
\usepackage{morefloats}
\usepackage{color}
\usepackage{xcolor}

\usepackage{enumitem}
\usepackage{hyperref}
\usepackage{multirow}
\usepackage{multicol}
\usepackage{ulem}

\newcommand{\kms}{km\,s$^{-1}$}

\newcommand{\msun}{M_\odot}
\newcommand{\bx}{}

\begin{document}

   \title{Massive dusty multiphase outflow in local merger shows no sign of slowing  on kiloparsec scales}

   \author{B. Hagedorn \inst{1}
          \and
          C. Cicone \inst{1} 
          \and
          M. Sarzi \inst{2}
          \and
          P. Severgnini \inst{3}
          \and
          C. Vignali \inst{4,5}
          }
      \institute{\inst{1}Institute of Theoretical Astrophysics, University of Oslo, P.O. Box 1029, Blindern, 0315 Oslo, Norway\\ 
      \inst{2}Armagh Observatory and Planetarium, College Hill, Armagh BT61 9DG, UK \\ 
      \inst{3}INAF - Osservatorio Astronomico di Brera, Via Brera 28, 20121 Milano, Italy \\ 
      \inst{4}Dipartimento di Fisica e Astronomia Augusto Righi, Università degli Studi di Bologna, via Gobetti 93/2, 40129 Bologna, Italy\\ 
      \inst{5}INAF – Osservatorio di Astrofisica e Scienza dello Spazio di Bologna, Via Gobetti 101, 40129 Bologna, Italy \\
      \email{bendix.hagedorn@astro.uio.no}
                      }
                
   \date{Received 23 July 2025 / Accepted 8 January 2026 }


\abstract{
We use ALMA CO(1-0) observations and VLT/MUSE rest-frame optical data of the ultraluminous infrared galaxy (ULIRG) IRAS20100-4156 at $z=0.1297$ to characterize its powerful outflow in multiple phases using tracers of cold molecular, ionized, and neutral atomic gas and dust as well.
Our analysis uses the correspondence with the stellar velocity field to split the complex emission line profiles of the CO(1-0) line into components in gravitational and non-gravitational motion.
We find a massive ($8\times10^{9}\msun$) molecular outflow containing about 40\% of the total molecular gas mass in the system.
The outflow shows a bi-conical morphology centered on the brightest galaxy in the merger, oriented along its minor axis and extending to $\sim5$\,kpc.
This outflow has a characteristic velocity of 170\,km/s, an outflow mass rate of 700\,$\msun/\mathrm{yr}$, a depletion time of 16\,Myr, and energetics consistent with star formation as a driver.
The neutral atomic and ionized gas phases traced by NaI absorption and H$\alpha$ emission show counterparts to the blueshifted cold molecular outflow but are only 15\% and 3\% as massive.
None of the three gas phases show any signs of slowing down over the extent at which we detected the outflow, suggesting an acceleration mechanism acting on the outflowing gas at kpc scales.
We also detect $3.5\times 10^7\msun$ of dust, traced by optical extinction in the MUSE data, in the blueshifted outflowing cold molecular gas.
The ionization state of the non-outflowing gas is consistent with star formation, while the outflowing component shows shock-like ionization.
We conclude that the multiphase outflow in IRAS20100-4156 originates in the southeast nucleus of the merger and is driven by the starburst activity there, with radiation pressure likely playing a significant role in its acceleration.
}

\keywords{galaxies: general -- galaxies: ISM -- galaxies: evolution}
\maketitle

\section{Introduction}
Galactic outflows are events in which gas from the interstellar medium (ISM) is accelerated to high velocities and possibly expelled from the galaxy, thereby altering its evolution.
Outflows can be multiphase and include cold and warm molecular gas, neutral gas, ionized gas, and dust.
The cold molecular gas phase, in particular, can carry a significant fraction of the total gas mass in some galaxies and play a role in regulating star formation \citep{murrayMaximumLuminosityGalaxies2005, feruglioQuasarFeedbackRevealed2010, ciconeMassiveMolecularOutflows2014}.
Some of the most extreme outflows have been observed in ultraluminous infrared galaxies (ULIRGs).
This class of galaxies is defined by an infrared (IR:\ $8-1000\,\mathrm{\mu m}$) luminosity in excess of $10^{12}\,L_\odot$ and, in the Local Universe, it consists entirely of objects undergoing major mergers.
Such mergers can trigger powerful starbursts and episodes of accretion and feedback from active galactic nuclei (AGNs; \citealt{ciconeMassiveMolecularOutflows2014}), which are thought to drive the observed outflows \citep[e.g.,][]{rupkeMULTIPHASESTRUCTUREPOWER2013,lampertiPhysicsULIRGsMUSE2022}.
The injection of energy into the interstellar medium (ISM) and possible expulsion of gas through outflows is thought to be a crucial step in the evolution of galaxies from blue star-forming systems to the red ``retired'' galaxies observed in our Universe \citep[e.g.,][]{erbFeedbackLowmassGalaxies2015}.
Understanding the role of outflows in this quenching process requires accurate measurements of the mass they contain and acquire from the galaxy for the purposes of star formation.
Spatially resolved spectroscopy is necessary to produce such measurements and constrain the relevant outflow properties \citep[e.g.,][]{muller-sanchezOUTFLOWSACTIVEGALACTIC2011,sotoEMISSIONLINESPECTRAMAJOR2012}.
Galactic outflows might be driven by star formation, through stellar winds and supernovae, or by AGNs through a variety of mechanisms \citep[see][for a review]{veilleuxGalacticWinds2005}, such as kinetic energy injection, radiation pressure, cosmic rays, or radio jets \citep[e.g.,][]{faucher-giguerePhysicsGalacticWinds2012,walchEnergyMomentumInput2015,thompsonDynamicsDustyRadiationpressuredriven2015}.
While many processes are theoretically capable of driving outflows,
detailed resolved outflow studies are necessary to constrain their structure and energetics which helps shed light on the relative importance of different driving mechanisms.
In ULIRGs, where both starburst and AGN activity are common and can be extremely intense, their relative importance in driving the ubiquitous outflows is unclear \citep[e.g.,][]{fluetschColdMolecularOutflows2018,lutzMolecularOutflowsLocal2020,lampertiPhysicsULIRGsMUSE2022}.
\begin{table}[tbp]
\caption{Relevant auxiliary data on IRAS20100-4156.}
    \centering
    \small
    \begin{tabular}{cccccc}
        \hline
        \hline\\[-8pt]
        $z_\mathrm{opt}$ & $D_L$ & $d_\mathrm{nuc,CO}$ & $\log(L\mathrm{_{IR}})$ & $\log(M_{*})$ & SFR  \\
        & [Mpc] & [kpc] & $[\rm L_{\odot}]$ & $[\rm M_{\odot}]$ &  $[\rm M_{\odot}~ yr^{-1}]$ \\
        (1) & (1) & (1) & (2) & (3) & (3) \\
        \hline\\[-8pt]
        0.1297 & 629 & 6.86 & 12.34 & 11.74 & 198 \\
         \hline
    \end{tabular}
    \caption*{Notes:  Numbers in parentheses indicate the source of the information: (1) this work; (2) \citet{ducSouthernUltraluminousInfrared1997}; and (3) \citet{gowardhanDualRoleStarbursts2018}}
    \label{tab:source_info}
\end{table}
The presence of cold dense gas in outflows presents its own puzzle, as the energetic processes driving the outflow are also capable of destroying molecules and dust \citep{KleinCloudCrushing1994}.
Entrainment of cold clouds in hot outflows \citep{gronkeGrowthEntrainmentCold2018} and in situ formation of cold gas via cooling in the outflow \citep{ZubovasGalaxyWideOutflows2014, ferraraFORMATIONMOLECULARCLUMPS2016} are possible explanations for this, but the  extent to which these processes are sufficient to explain the observed massive cold molecular outflows remains unknown \citep[see][for a review]{veilleuxCoolOutflowsGalaxies2020}.
Due to their multiphase nature, capturing a complete picture of galactic outflows requires multiwavelength observations \citep{rupkeMULTIPHASESTRUCTUREPOWER2013,ciconeLargelyUnconstrainedMultiphase2018}, accounting for all relevant gas phases and possible dust components.
This is especially crucial when comparing to theoretical predictions from hydrodynamical cosmological simulations, which generally are not equipped to model all gas phases separately; rather, they rely on subgrid prescriptions to implement multiphase gas \citep{somervillePhysicalModelsGalaxy2015}.
In this work, we aim to leverage the already necessary multiwavelength data further to improve the outflow identification process.
The most widely used techniques for identifying outflows rely on analyzing the kinematics of the gas through observations of spectral lines in emission or absorption \citep[e.g.,][]{rupkeINTEGRALFIELDSPECTROSCOPY2011,ciconeMassiveMolecularOutflows2014}.
This can be challenging in kinematically complex systems, such as ULIRGs, as there are a number of degeneracies between outflows, tidal features, merger parts, and turbulent motion, all of which can produce gas at high velocities \citep{rupkeMULTIPHASESTRUCTUREPOWER2013}.
The nearby ULIRG IRAS20100-4156 has been studied primarily as part of larger samples of local ULIRGs \citep{spoonDIAGNOSTICSAGNDRIVENMOLECULAR2013, fluetschColdMolecularOutflows2018, fluetschPropertiesMultiphaseOutflows2021, pernaPhysicsULIRGsMUSE2021, lampertiPhysicsULIRGsMUSE2022}, along with its powerful outflow characterized based on integrated spectra.
Previous resolved studies of the cold molecular outflow in IRAS20100-4156 have relied on separating only the very high velocity gas in the object to identify the outflow \citep{gowardhanDualRoleStarbursts2018, lampertiPhysicsULIRGsMUSE2022}. This approach  offers the potential to underestimate the total outflow mass by excluding outflowing gas at low velocities, as noted by \citet{lampertiPhysicsULIRGsMUSE2022}.
This technique, developed for the analysis of isolated star-forming galaxies with gas dynamics dominated by regular rotating motion, might not be sufficient to capture the complexity of a merging system such as IRAS20100-4156.
In this paper, we test a physically motivated outflow identification method, combining submillimeter (submm) and optical spectroscopy. We used it to derive the physical properties of the cold molecular outflow in IRAS20100-4156.
In this method, the spectral decomposition of the CO(1-0) emission line is based on stellar kinematics derived from optical spectroscopy at similar spatial resolution.
A similar technique was recently applied to near-IR (NIR) observations of the merger NGC6240 \citep{carlsenOutflowingShockedGas2025}.
We further leveraged the multiwavelength data to compare the kinematics of the outflow between different gas phases resolved at the same scales and to identify a dust component in the outflow, which has not  been done for this object in prior studies.
The paper is structured as follows.
In Sect.~\ref{sec:data}, we describe the target and the data used in this work.
Section~\ref{sec:methods} describes the methods used in fitting the spectra and identifying the outflows.
In Sect.~\ref{sec:results}, we derive outflow properties and, in Sect.~\ref{sec:discussion}, we discuss their implications for the origin and driving mechanism of the outflow and the evolution of the galaxy.
In Sect.~\ref{sec:conclusion} we offer a brief summary and conclude with the main findings of this work.
Throughout this paper, we assume a flat $\mathrm{\Lambda CDM}$ cosmology with parameters from the \citet{planckcollaborationPlanck2018Results2020} for the luminosity distance calculations.

\section{Data}
\label{sec:data}
\begin{figure}[tbp]
    \centering
    \includegraphics[clip=true, trim=2cm 0.7cm 1.5cm 0.1cm,width=0.8\linewidth]{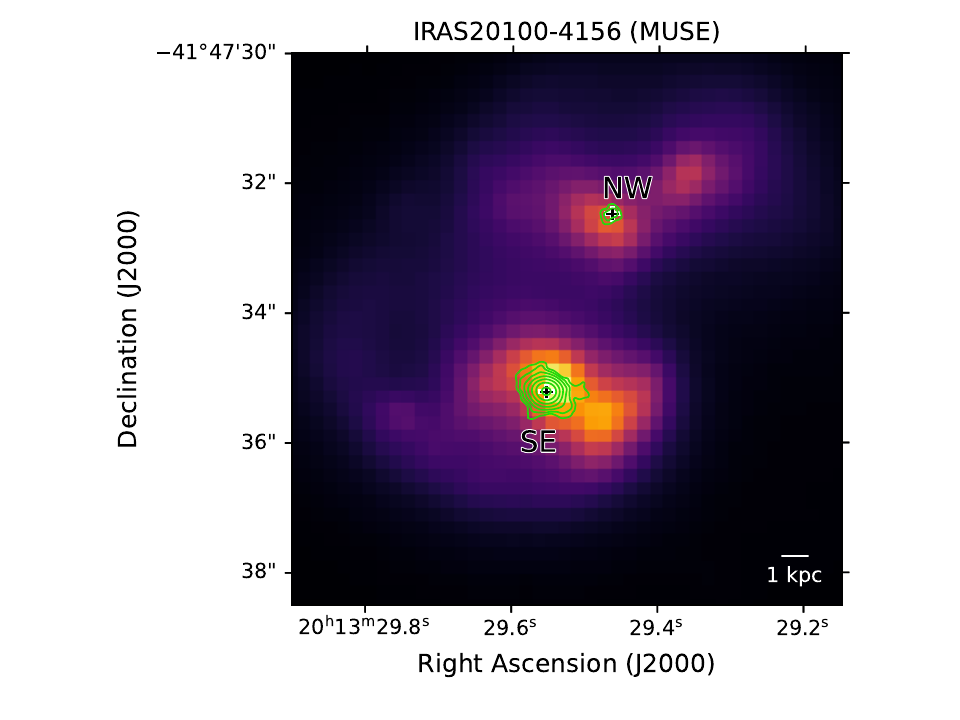}
    \caption{Integrated intensity map of IRAS20100-4156 produced from the MUSE cube. The black-and-white crosses show the positions of the two main merger components determined from the high-resolution ALMA observations of CO(3-2), while the green contours show the zeroth moment of the same CO(3-2) data. The first two contour levels correspond to three and five times the noise level of the moment map, respectively, and every following step is doubled.}
    \label{fig:muse_overview}
\end{figure}

\subsection{IRAS20100-4156}
\label{sec:iras20100-4156}
IRAS20100-4156 is a merger consisting of two prominent components with a projected separation of $\sim2.87"$ corresponding to $\sim6.86$\,kpc at a redshift of 0.1297.
Using HST imaging  \citet{fluetschPropertiesMultiphaseOutflows2021} classified the system as pre-merger with a close binary (IIIb), adopting the classification scheme from \citet{veilleuxOpticalNearInfraredImaging2002}.
There are no direct observations of an AGN in IRAS20100-4156 and indirect evidence is inconclusive.
Based on optical spectroscopy, IRAS20100-4156 has been classified as variously as starburst (SB) dominated \citep{ducSouthernUltraluminousInfrared1997} and LINER-like \citep{pernaPhysicsULIRGsMUSE2021} using different classification schemes and apertures for the extraction of integrated spectra.
Similarly, its X-ray spectral properties have been reported as intermediate between an SB and AGN system \citep{franceschiniXMMNewtonHardXray2003}.
Mid-infrared (MIR) observations of IRAS20100-4156 show no significant emission in the 14.32$\mu$m [NeV] line \citep{spoonDIAGNOSTICSAGNDRIVENMOLECULAR2013}, a tracer of AGN activity, providing no clear evidence of an AGN either.
\citet{lampertiPhysicsULIRGsMUSE2022} reported the presence of a compact obscured nucleus, based on ratios of equivalent widths (EWs) of different polycyclic aromatic hydrocarbon (PAH) features.
However, it is not clear whether the presence of such compact obscured nucleus is evidence of an AGN or can be explained by starburst activity \citep{veilleuxSPITZERQUASARULIRG2009, aaltoProbingHighlyObscured2015}.
Finally, \citet{SpoonMIRgalaxyClassification2007} found IRAS20100-4156 to exhibit a combination of silicate absorption and PAH emission that places it between a typical obscured nucleus and typical starburst galaxy in their MIR classification scheme.
The main properties of the target are listed in Table~\ref{tab:source_info}.

\subsection{ALMA observations}
IRAS20100-4156 was observed with  Atacama Large Millimeter/submillimeter Array (ALMA) in band 3 (program ID: 2013.1.00659.S, PI: H. Spoon), which covers the CO(1-0) emission line at a rest-frame frequency of 115.271\,GHz.
These data were obtained from the ALMA ESO archive and reduced using CASA version 6.1.0-118.
The dataset consists of five separate observations taken between  June 9 and 15, 2014, using between 33 and 35 ALMA 12m antennae in the C34-4 configuration.
This results in an angular resolution of 1$''$, a maximum recoverable scale of 11.3$''$, and a field of view (FoV) of 57.6$''$.
The estimated precipitable water vapor (PWV) for these observations was 0.716\,mm and the estimated line sensitivity (for a line width of 10\,km/s)
0.496\,mJy/beam.
Each of the subsets where continuum subtracted using \texttt{uvconsub} and combined into a single measurement set using \texttt{concat}.
The measurement set was cleaned with \texttt{tclean}, producing a data cube with a pixel scale of 0.1$''$ and a spectral resolution of 7.2\,km/s.
This cube was then corrected for the primary beam using \texttt{impbcor}.
The final data product has an RMS of 0.11\,mJy and a median beam size of $1.27''\times1.05''$ (major and minor axis).
We show maps of the first three moments for this data cube in Appendix~\ref{app:CO_moment_maps}.
In addition, we used high-angular-resolution ($\sim0.16''$) ALMA band 7 observations (program ID: 2018.1.00888.S, PI: A. Gowardhan) of the CO(3-2) line in order to define the centers of the two main merger components (see Fig.~\ref{fig:muse_overview}). We label them as SE and NW nucleus based on their relative positions.

\subsection{MUSE observations}
\label{sec:data:muse}
We obtained VLT/MUSE  (Multi Unit Spectroscopic Explorer) data for IRAS20100-4156 from the ESO archive (program ID: 0103.B-0391, PI: S. Arribas). 
The data were reduced with the MUSE EsoReflex \citep{freudlingAutomatedDataReduction2013} pipeline recipes (muse – 2.6.2) and the residual sky contamination was subtracted using the using the Zurich Atmosphere Purge (\texttt{ZAP}) software package \citep{SotoZAP2016}.
These observations were performed with the assistance of adaptive optics, resulting in an angular resolution of 0.6$''$ and a 1.56$'$ FoV.
Figure~\ref{fig:muse_overview} shows the subsection of the FoV used in our analysis.
The MUSE data cover a spectral range from 4750\,\AA\ to 9350\,\AA\ and have a mean resolution of 2.65\,\AA, corresponding to $\sim120\,$km/s.
The spectrum is sampled with a bin size of 1.25\,\AA, corresponding to $\sim60\,$km/s.
In this work, we focus on the [OIII]$\lambda5007$ and H$\alpha$ emission lines to characterize the ionized phase of the outflow.
The NaI$\lambda\lambda$5890,96 doublet absorption feature was also covered by the data, as it is redshifted out of the 5800\AA-5970\AA~window that is blocked to avoid contamination from the AO sodium laser; thus, we used it as a tracer for the neutral atomic phase in the outflow. To fix the astrometric discrepancies between the ALMA and MUSE data, an astrometry correction was applied to the MUSE cube based on four stars with known positions from the Gaia catalog and HST imaging using the Starlink-Gaia software \citep{Draper2014ascl.soft03024D}.

\section{Methods}
\label{sec:methods}
The complex kinematic structure of IRAS20100-4156 results in complicated CO(1-0) emission line profiles in the high spectral resolution ALMA data, featuring multiple peaks and asymmetries in parts of the galaxy.
We aim to decompose these lines into emission arising from outflowing gas and gas in gravitationally induced motion.
In a major merger such as IRAS20100-4156, we would expect gravitationally induced motions to deviate from those of an unperturbed rotating disk due to the gravitational interaction between merging galaxies.
To account for this added kinematic complexity in the non-outflowing gas, we used the stellar kinematics as a tracer for gravitationally induced motions in the molecular gas, assuming that stars and molecular gas are sufficiently well coupled on the scales resolved by ALMA and MUSE in IRAS20100-4156 to make this a good approximation.

\subsection{Fitting for stellar kinematics}
\label{sec:methods:gandalf_stars}
\begin{figure}[tbp]
    \centering
    \includegraphics[clip=true, trim=0.3cm 0.1cm 0.6cm 0.1cm,width=0.42\textwidth]{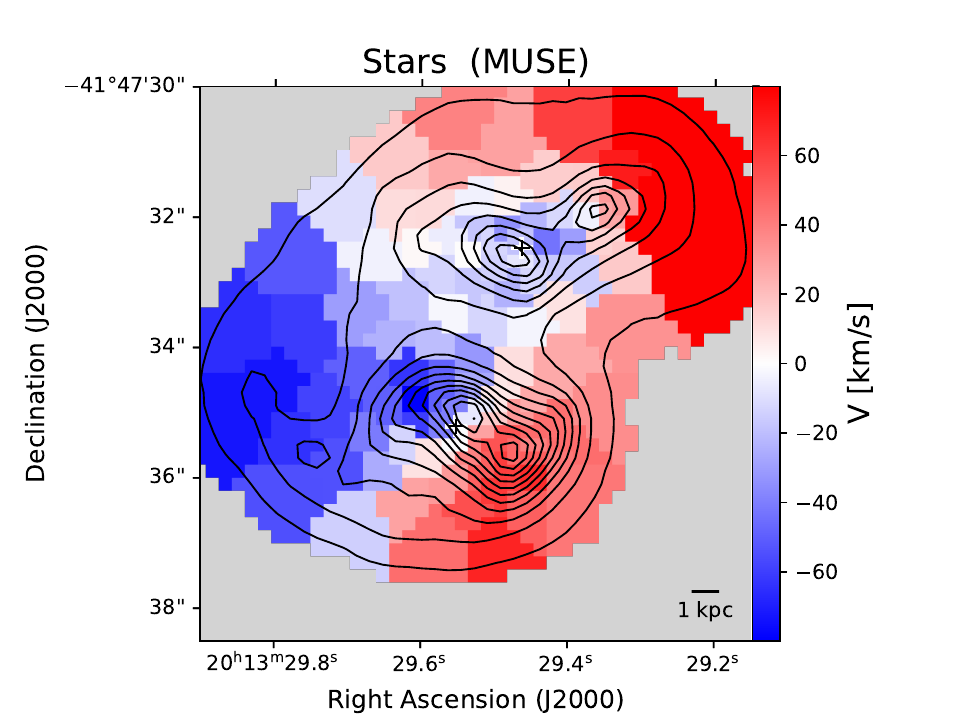}
    \caption{Velocity map of the stellar population in Voronoi bins with a target S/N of 40 in the continuum. The black contours show the stellar continuum emission integrated over the entire spectral range of the fit.}
    \label{fig:vorbin_vel_maps_stars}
\end{figure}

We use the rest-frame optical MUSE spectroscopy to constrain stellar kinematics in IRAS20100-4156.
The cube was Voronoi-binned with a target signal-to-noise ratio (S/N) of 40 for each bin to ensure we have a sufficiently high signal in the stellar continuum across the galaxy to fit stellar absorption line features, while retaining sufficient spatial information to capture the morphology of the outflow.
We used \texttt{pPXF} \citep{cappellariParametricRecoveryLineofSight2004} to fit stellar population templates and kinematics to the Voronoi-binned data.
This results in a central velocity and line-of-sight velocity dispersion (LoSVD) for each bin.
Figure~\ref{fig:vorbin_vel_maps_stars} shows a map of the central velocity in each bin.
Details on the fitting procedure can be found in Appendix~\ref{app:gandalf}.

\subsection{Outflow identification in the cold molecular phase}
\label{sec:methods:outflow_identification}
\begin{figure}
    \centering
    \includegraphics[clip=true, trim=0.6cm 0 0 0,width=0.49\textwidth]{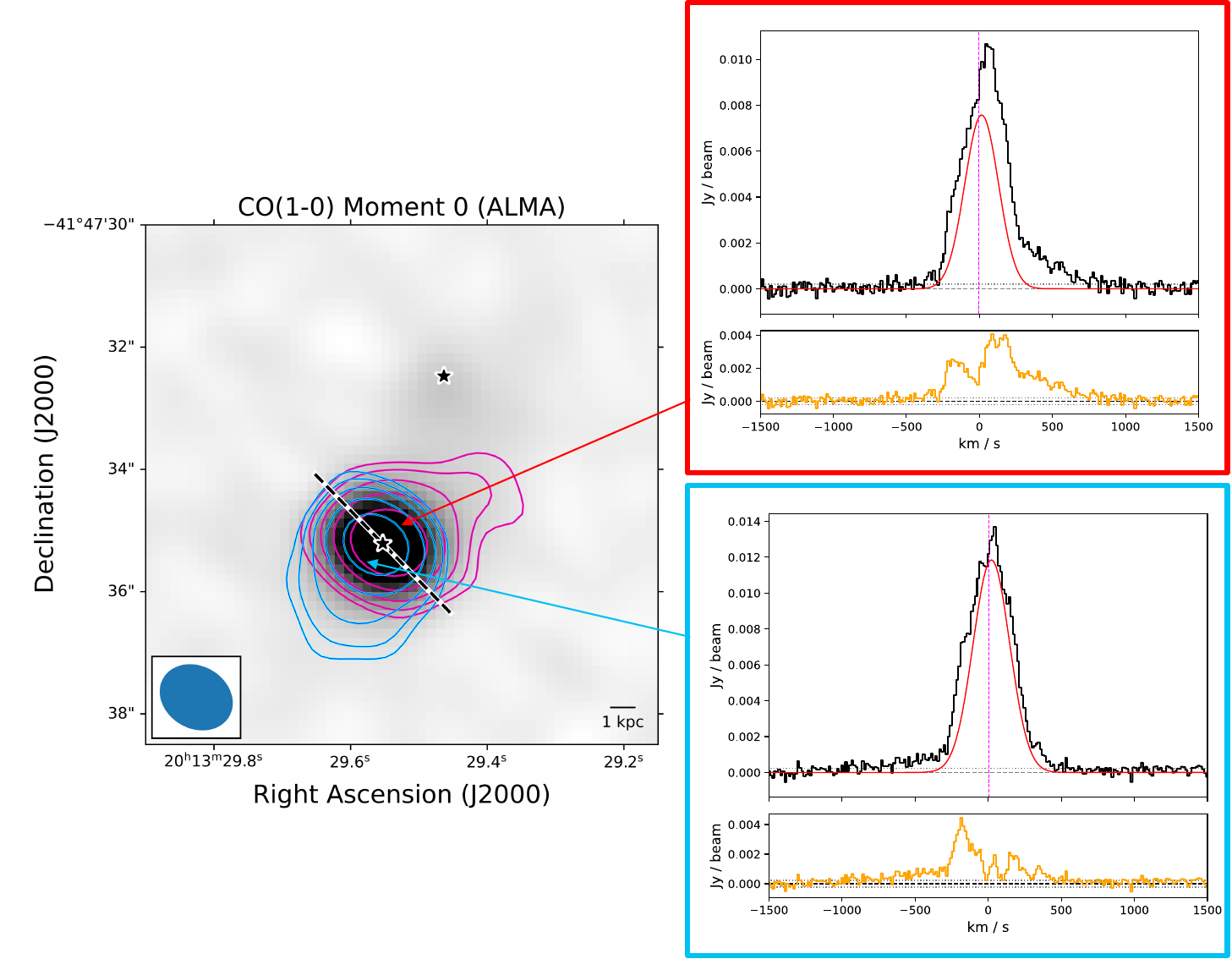}
    \caption{Moment 0 map of CO(1-0) emission. The color map is cut at half the maximum to highlight fainter structure, saturating the SE nucleus. The black-and-white star markers show the positions of the CO(3-2) emission peaks in the SE and NW nuclei. The black-and-white dashed line shows the estimated orientation of the major axis in the SE disk. The cyan-and-blue (magenta-and-red) contours show the integrated intensity of emission from gas with velocities below -150\,km/s (above 150\,km/s) relative to the systemic velocity. The first two contour levels correspond to the 3$\sigma$ and 5$\sigma$ significance levels and every following step is doubled. The cyan and red framed panels show spectra of the CO(1-0) line (black) extracted from two positions along the minor axis. The vertical magenta dashed line in the extracted spectra shows the line-of-sight velocity of the stellar component at that position. The red line profile corresponds to the fitted quiescent gas components. The bottom panel shows the spectrum of the outflowing gas in orange. In both panels, dashed horizontal lines show the zero-level and dotted horizontal lines the 1$\sigma$ noise level.}
    \label{fig:co10_vorbin_fit_rot_examples}
\end{figure}
The first step in our outflow identification procedure is Voronoi binning the CO(1-0) data cube using the \texttt{vorbin} python package \citep{cappellariAdaptiveSpatialBinning2003} with a target S/N of 50.
This S/N value is comparable to the highest S/N in individual pixels, resulting in a relatively uniform S/N across all bins.
As a consequence, the highest S/N pixels around the SE nucleus remain unbinned.
This results in 162 bins, covering the SE and NW nuclei in the merger.
A map of the binning can be found in Appendix~\ref{app:co_line_fitting}.
For each Voronoi bin, we then determined the average stellar kinematics using the results from the \texttt{pPXF} fit to the Voronoi-binned MUSE data.
To this end, we first re-projected the Voronoi binned stellar velocity field into the coordinate system of the ALMA data.
Then we took the mean stellar velocity of all pixels in the ALMA Voronoi bins (which are different from the MUSE Voronoi bins).
Next, we iteratively fit up to five components to the complex CO(1-0) line profiles using the \texttt{pyspeckit} package \citep{ginsburgPyspeckitSpectroscopicAnalysis2022, pyspeckitSoftware}, starting with a single component initialized with the stellar central velocity and velocity dispersion.
This component is used to model gas in gravitationally induced motion, such as disk rotation or tidal features, which (unlike outflowing gas) is likely to have a counterpart in the stellar component.
Hereafter, we call this the quiescent component of the gas, or quiescent gas.
In the fitting procedure, this first component's central velocity and velocity dispersion were constrained to avoid deviating from the stellar kinematics by more than three times the uncertainties on stellar central velocity and velocity dispersion obtained in the \texttt{pPXF} fit.
Up to four additional Gaussian components were then added to the fit in an iterative process.
We found this to be the lowest number of additional components sufficient to model prominent features in the complex line profiles throughout in the galaxy. 
This addition of extra components allows the primary component to capture only the quiescent gas.
Each of the additional components was left free to vary in amplitude, central velocity, and velocity dispersion.
The results of every iteration were saved and the best fitting solution was chosen by computing the Akaike information criterion (AIC) for each and choosing the fit with the lowest AIC.
We performed a visual check of the fit for each bin to ensure its quality.
Examples of the spectral fitting can be found in Appendix~\ref{app:co_line_fitting}.
After fitting for the quiescent gas component, we subtracted this component from the spectrum in each bin.
This left us with emission from gas in nongravitationally induced motion, such as outflows.
In the following, we call this the outflowing component or outflowing gas.
Figure~\ref{fig:co10_vorbin_fit_rot_examples} shows examples of extracted spectra, best-fit quiescent components, and the resulting spectra of outflowing gas emission.
\begin{figure*}[tbp]
    \centering
    \includegraphics[clip=true, trim=0.1cm 0.1cm 0.1cm 0.05cm,width=0.99\textwidth]{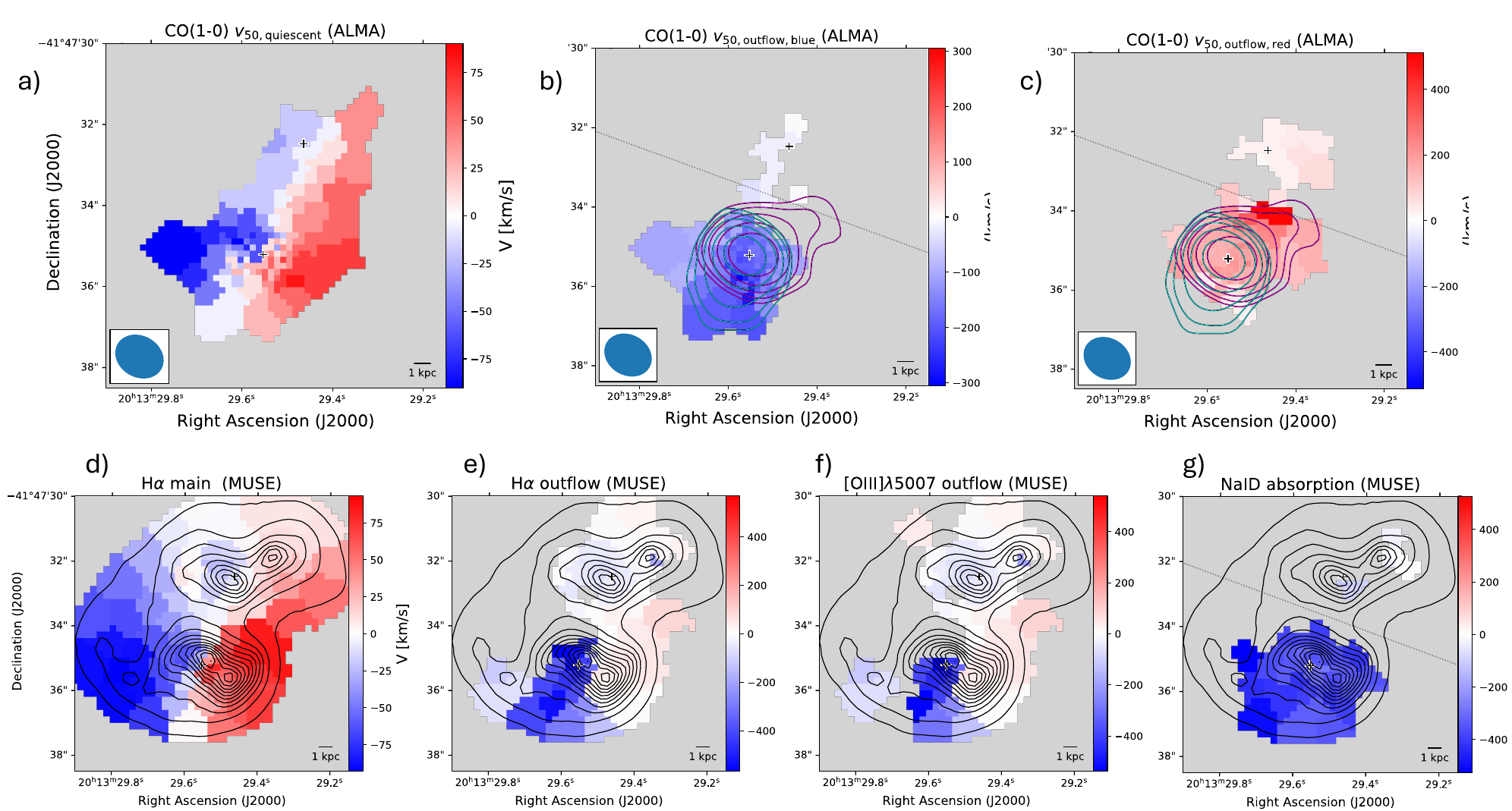}
    \caption{{\bx Velocity maps of quiescent and outflowing gas in the cold molecular, ionized and neutral gas phases.
    Top row: Maps of $v_{50}$ for the quiescent CO(1-0) emission (a), blueshifted outflowing (b), and redshifted outflowing (c) gas. The cyan-and-blue (magenta-and-red) contours show the integrated intensity of emission from gas with velocities below -150\,km/s (above 150\,km/s) relative to the systemic velocity. The first two contour levels correspond to the 3$\sigma$ and 5$\sigma$ significance levels and every following step is doubled. The dotted lines in the outflow panels show the borders according to which bins are associated with the SE and NW nuclei when calculating total outflow properties.
    Bottom row: Velocity maps based on MUSE data. Quiescent ionized gas traced by H$\alpha$ (d), and outflowing ionized gas traced by H$\alpha$ (e) and by [OIII]$\lambda$5007 emission (f).
    Panel (g) shows the best-fit velocity map of the neutral outflowing gas traced by NaID absorption. The black contours show the stellar continuum emission integrated over the spectral range. We do not show the quiescent component traced by [OIII]$\lambda$5007, as kinematics are tied between this line and H$\alpha$ in the fit and the quiescent component is detected in every bin, so that this map would be identical to that of quiescent H$\alpha$.}
    }
    \label{fig:all_phases_vorbin_vel_maps}
\end{figure*}

\subsection{Outflow identification in the ionized phase}
\label{sec:methods:ion_outflow_identification}
We used the H$\alpha$ and [OIII]$\lambda5007$ lines to derive outflow properties for the ionized gas.
Due to the blending of the H$\alpha$ line with the [NII] doublet, especially in their associated broad outflow features, we were not able to apply the same technique for outflow identification here as for the CO(1-0) line.
Instead, we fit the optical emission lines with up to two Gaussian components as part of the full-spectrum fit with \texttt{GandALF} \citep{sarziSAURONProjectIntegralfield2006, sarziFornax3DProjectOverall2018}, which is described in more detail in Appendix~\ref{app:gandalf}.
We assumed that the primary, narrower, emission line component from the \texttt{GandALF} fit traces gas in gravitational motion, such as rotation and tidal features.
We ensured this approximation works reasonably well by comparing, in each bin, the resulting central velocity and velocity dispersion of the narrow component to that of the stars: 
we found no major deviations between these quantities, confirming that the narrow component traces gas coupled to the stars; hence, it is "quiescent" according to our definition.
The secondary component was then taken to trace outflowing gas, allowing us to obtain outflow velocities and luminosities for each Voronoi bin.
Due to the much lower spectral resolution of MUSE ($R\sim1770 - 3590$) compared to ALMA, the line profiles of the optical emission lines are less complex than those of the CO(1-0) line and this approach can successfully trace outflowing gas, as we describe in Sect.~\ref{sec:results:ionized_outflow}.
We applied the same approach to the [OIII] and other strong emission lines ([OI]$\lambda6300$, [NII]$\lambda6484$, and [SII]$\lambda\lambda6717,31$).
This allowed for the use of emission-line based diagnostics to determine the ionization state and electron density of the outflowing gas (see Sect.~\ref{sec:results:bpt_met} and Appendix~\ref{app:electron_density}).

\subsection{Outflow identification in the neutral atomic phase}
\label{sec:methods:neutral_outflow_identification}
The NaI$\lambda\lambda$5890,96 doublet (hereafter, simply the NaI doublet) absorption feature is present both in the stellar component and as a nebular absorption line.
The nebular absorption feature traces gas between the observer and the galaxy, illuminated by continuum emission from the galaxy itself.
With its low ionization energy (5.14\,eV), NaI is a good tracer of neutral hydrogen.
This combination makes blueshifted nebular NaI doublet absorption a direct and unambiguous tracer of outflowing gas in the neutral atomic phase.
We fit the NaI doublet absorption feature after accounting for the stellar continuum and HeI$\lambda$5876 emission as derived from the \texttt{GandALF} fit, using the interstellar absorption line model based on \citet{rupkeOutflowsInfraredLuminousStarbursts2005} and \citet{satoAEGISNATUREHOST2009}. Finally, we used the results to derive the outflow properties of the neutral atomic gas phase.
Details on the fitting procedure and the derivation of outflow properties can be found in Appendix~\ref{app:Na_fitting}.

\section{Results}
\label{sec:results}

\subsection{Cold molecular gas}
\label{sec:results:mol_outflow}

\begin{table}[tbp]
    \small
        \caption{Molecular gas properties by kinematic component.}
    \centering
    \begin{tabular}{lcccc}
    \hline
    \hline\\[-8pt]
     Component & $S_\mathrm{CO}$ & $L'_\mathrm{CO}$ & $M_\mathrm{mol}$ \\[3pt]
     & $[\rm Jy~\frac{km}{s}]$ & $[\mathrm{10^9K\frac{km}{s}pc^2}]$ & $[10^9\,M_\odot]$\\
     \hline
quiescent          & $7.9\pm 0.06$ & $6.75\pm 0.05$ & $11\pm 3.4$ \\
blue OF SE & $2.3\pm 0.10$ & $1.98\pm 0.09$ & $3.4\pm 1.0$ \\
red OF SE  & $2.9\pm 0.05$ & $2.52\pm 0.04$ & $4.3\pm 1.3$ \\
total OF SE& $5.3\pm 0.6$ & $4.51\pm 0.10$ & $8\pm 2.3$ \\
blue OF NW & $0.030\pm 0.008$ & $0.029\pm 0.007$ & $0.049\pm 0.018$ \\
red OF NW  & $0.16\pm 0.020$ & $0.139\pm 0.017$ & $0.24\pm 0.08$ \\
total OF NW& $0.20\pm 0.021$ & $0.168\pm 0.018$ & $0.29\pm 0.09$ \\
    \hline
    \end{tabular}
\caption*{Notes: Integrated CO total line fluxes, luminosities and the corresponding molecular gas masses listed are summed over all Voronoi bins for each of the kinematic cold molecular gas components.}
    \label{tab:molgas_totals}
\end{table}
We split the outflowing gas spectrum, identified as described in Sect.~\ref{sec:methods:outflow_identification}, into a blueshifted and a redshifted half (relative to the redshift of the system obtained from the fit to the MUSE data) for each bin.
For each half we compute the flux and 50th percentile velocity of all blue or redshifted emission respectively in the bin by integrating until the flux density drops below the noise level, calculated as the standard deviation of the spectrum where there is no line emission.
We use $v_\mathrm{50,blue}$ ($v_\mathrm{50,red}$) as the characteristic velocity of the approaching blueshifted (receding redshifted) outflowing cold molecular gas in the bin.
In the following analysis, we include only the bins in which the maximum of the blueshifted (redshifted) outflowing spectrum exceeds three times the standard deviation of the flux density measured away from the line emission in at least three spectral channels.
This ensures we exclude bins that do not have significant emission from outflowing gas, which would lead to spurious results in the characteristic outflow velocity in the following calculations.
Panels (b) and (c) in Fig.~\ref{fig:all_phases_vorbin_vel_maps} show maps of the characteristic velocities of the detected blue- and redshifted outflowing gas components alongside maps for the quiescent gas molecular gas component (panel a) and the ionized ant neutral atomic phases based on MUSE data (panels d-g).
We also show the integrated intensity of emission originating from gas at high velocities in the complete ALMA cube (no quiescent component subtracted) overlaid as blue and magenta contours in Fig.~\ref{fig:all_phases_vorbin_vel_maps}.
We set the threshold for what constitutes high-velocity gas at 150\,km/s offset from the systemic velocity.
This is about a factor of two higher than the maximum central velocity of stars across the bins (see Fig.~\ref{fig:vorbin_vel_maps_stars}) and also lies above the typical velocity dispersion of stars in the bins. 
We can therefore safely assume that emission from beyond this threshold traces mainly non-gravitational motions in the molecular gas.
Around the SE nucleus, the morphology of the blue- ($v < -150$\,km/s) and redshifted ($v > 150$\,km/s) high-velocity gas components is very similar to that of the outflowing gas selected from the fit, with the redshifted emission stretching toward the NW and the blueshifted emission toward the SE along the minor axis (see Fig.~\ref{fig:all_phases_vorbin_vel_maps}).
This clear correlation between the high-velocity gas component and the gas we identify as outflowing after subtracting the quiescent component serves as a check to confirm that our emission-line decomposition indeed identifies gas in outflow, rather than incoherent fit residuals or differential rotation between molecular gas and stars.
Around the NW nucleus, the outflowing gas component lacks the same clear morphology and there is no detection of high-velocity gas to compare it with.
The presence of an outflow in the NW is therefore uncertain, noting that its contribution to the overall outflow flux and energetics would be insignificant compared to the SE outflow (see Table~\ref{tab:molgas_totals}).
We assigned each bin to one of the two main merger components, based on their distance.
The dividing line is shown in panels (b), (c), and (g) of Fig.~\ref{fig:all_phases_vorbin_vel_maps} and was chosen based on the morphology of the outflowing gas.
We used this division to calculate total outflow fluxes and derived properties for each component (denoted SE and NW outflow, respectively) separately.
For each component (blue- and redshifted) in every bin, we calculated the CO(1-0) line luminosity $L'_\mathrm{CO}$ following \citet{solomonMolecularInterstellarMedium1997}.
From the line luminosities the corresponding molecular gas masses are then calculated using the CO-to-H$_2$ conversion factor $\alpha_\mathrm{CO}$ with a value of $1.7\,M_\odot\,\mathrm{(K\,km/s\,pc^2)}^{-1}$ from a recent study of ULRIGs \citep{montoyaarroyaveSensitiveAPEXALMA2023}.
This value has been calibrated for the bulk of the molecular gas in a sample of local ULIRGs similar to IRAS20100-4156, using sensitive observations of CO and [CI] emission lines. 
Since no additional tracers of molecular gas, such as [CI], have been observed in IRAS20100-4156 directly, this represents the most accurate calibration of $\alpha_\mathrm{CO}$ available for this target.
The SE molecular outflow is extremely massive with a total mass of $8\times10^9\msun$ split into blue- and redshifted components.
This accounts for about 40\% of the total molecular gas mass in the system ($19\times10^9\msun$ for quiescent and outflowing gas combined).
The results for all components are listed in Table~\ref{tab:molgas_totals}.

\subsection{Ionized gas tracers}
\label{sec:results:ionized_outflow}

\begin{table}[]
    \small
    \caption{Ionized gas properties by fit component and tracer across Voronoi bins.}
    \centering
    \begin{tabular}{lccc}
        \hline
        \hline\\[-8pt]
         Component & $F_\mathrm{line}$ & $L_\mathrm{line}$ & $M_\mathrm{gas}$\\[3pt] 
         & $[10^{-15}\mathrm{\frac{erg}{s\cdot cm^2}}]$ & $[10^9 L_\odot]$ & $[10^7 M_\odot]$\\
         \hline
H$\alpha$ main              & $367\pm 7.9$ & $4.54\pm 0.098$ & $80\pm 24$ \\
H$\alpha$ wing SE           & $72.0\pm 1.7$ & $0.89\pm 0.021$ & $9\pm 3.0$ \\
H$\alpha$ wing NW           & $33\pm 2.8$ & $0.41\pm 0.035$ & $8\pm 3.0$ \\
$\mathrm{[OIII]}$ main      & $123\pm 1.2$ & $1.519\pm 0.015$ & $5.1\pm 1.3$ \\
$\mathrm{[OIII]}$ wing SE   & $23.1\pm 0.09$ & $0.285\pm 0.0011$ & $0.5\pm 0.30$ \\
$\mathrm{[OIII]}$ wing NW   & $12\pm 0.25$ & $0.15\pm 0.0031$ & $0.4\pm 0.29$ \\
         \hline
    \end{tabular}
    \label{tab:iongas_masses}
\end{table}

From the \texttt{Gandalf} fit to the MUSE data (see Appendix~\ref{app:gandalf} for details), we can directly obtain characteristic outflow velocities for each Voronoi bin as the central velocity of the Gaussian emission line component.
Panels (d), (e), and (f) of Fig.~\ref{fig:all_phases_vorbin_vel_maps} show velocity maps for the quiescent and outflowing components of the ionized gas phase respectively.
The quiescent ionized gas shows a similar rotation-like pattern to the stellar component (compare Figs.~\ref{fig:vorbin_vel_maps_stars} and~\ref{fig:all_phases_vorbin_vel_maps}), indicating that the primary component used in the fit indeed traces quiescent gas.
The outflowing component shows a prominent blueshifted feature to the SE of the SE nucleus, similar to that seen in CO(1-0) but less extended.
The shared morphology with the blueshifted cold molecular outflow suggests that this feature corresponds to the ionized phase in a multiphase outflow originating from the SE nucleus.
Aside from this feature across the rest of the galaxy, the secondary component shows much lower velocities and no other structures as clearly suggestive of outflowing gas.
We explore the possibility that the second component may not trace an outflow at all in these bins further in Sect.~\ref{sec:discussion:outflow_velocity_structure}.
We obtained the total mass contained in the ionized phase of the outflow by adding the contributions of all bins in which an outflow signature is detected.
We set our threshold for a detection at $A_\mathrm{G}/N_\mathrm{G} > 3$, where $A_\mathrm{G}$ and  is the amplitude of the corresponding emission line fit component and $N_\mathrm{G}$ is the noise level determined in the \texttt{GandALF} fit.
The conversion of luminosities of H$\alpha$ and [OIII]$\lambda$5007 into gas masses depends on the electron density in the outflowing gas.
We calculated the electron densities for each bin from the ratio between the [SII]$\lambda$6716 and [SII]$\lambda$6731 emission lines $R=f_\mathrm{[SII]6716}/f_\mathrm{[SII]6731}$ following \citet{sandersMOSDEFSURVEYELECTRON2016}. Hereafter, we refer to these lines as the [SII] doublet).%
\footnote{{\bx An alternative method for computing electron densities in the case of galactic outflows has been introduced by \cite{baronDiscoveringAGNdrivenWinds2019}. That method applies to AGN-driven outflows and as long as the AGN is the main source of ionization. Since our outflow region appears dominated by LINER-like emission most likely powered by shocks we do not consider it applicable to our case.}}
The IDL version of \texttt{lmfit} used in the \texttt{GandALF} full-spectrum fit does not allow R to be constrained to an interval corresponding to theoretical limits on the line ratio.
We therefore used the \texttt{lmfit} python package to re-fit the amplitudes of both main and wing components for both lines, while keeping the line kinematics fixed to the results from the \texttt{GandALF} full-spectrum fit.
The Python implementation of \texttt{lmfit} allowed us to constrain $R$ to lie within 0.44 and 1.45, corresponding to the theoretical extremes which the line ratio approaches asymptotically at very high ($n_\mathrm{e}>10^5\,\mathrm{cm^{-3}}$) and very low ($n_\mathrm{e}<1\,\mathrm{cm^{-3}}$) electron densities respectively  \citep[see Fig. 1 in][]{sandersMOSDEFSURVEYELECTRON2016}.
Even within these extremes, the [SII] doublet line ratio is only sensitive to the electron density  in the range of $50-5000\,\mathrm{cm^{-3}}$ due to its asymptotic behavior.
We accounted for this by treating electron densities below $50\,\mathrm{cm^{-3}}$ as upper limits with value $50\,\mathrm{cm^{-3}}$ in our calculations to avoid the final results being dominated by bins with highly uncertain low electron densities.
As a result, the global outflow masses we obtained should be treated as the lower limits for the actual outflow masses.
The masses were inferred from H$\alpha$ luminosities for each bin following \citet{fluetschPropertiesMultiphaseOutflows2021}:
\begin{equation}
    M_\mathrm{ion,H\alpha}=6.1 \times 10^8 \left(\frac{L_\mathrm{H\alpha}}{10^{44}\,\mathrm{\frac{erg}{s}}}\right) \left(\frac{500\,\mathrm{\frac{1}{cm^{3}}}}{n_\mathrm{e}}\right)\,M_\odot,
\end{equation}
where $n_\mathrm{e}$ is the electron density in the outflowing gas.
We calculated the independent outflow masses based on the [OIII]$\lambda$5007 line according to \citet{carnianiIonisedOutflows242015}:
\begin{equation}
\label{eq:ion_mass}
    M_\mathrm{ion,[OIII]}=8 \times 10^7 \left(\frac{L_\mathrm{[OIII]}}{10^{44}\,\mathrm{\frac{erg}{s}}}\right) \left(\frac{500\,\mathrm{\frac{1}{cm^{3}}}}{n_\mathrm{e}}\right) \left(\frac{C}{10^\mathrm{[O/H]-[O/H]_\odot}}\right)\,M_\odot,
\end{equation}
with an additional dependence on the ``condensation factor' of' $C=\langle n_\mathrm{e} \rangle^2/\langle n_\mathrm{e}^2 \rangle$ and the gas-phase metallicity relative to the solar value expressed as $\mathrm{[O/H]-[O/H]_\odot}$.
Each bin was then associated with either the SE or NW nucleus using the same criterion as for the cold molecular gas.
We then took the sum over all Voronoi bins associated with each nucleus and in which the respective gas tracer is detected to obtain the total ionized outflow masses for the SE and NW nuclei separately.
The results are listed in Table~\ref{tab:iongas_masses}.
The ionized outflow mass based on [OIII]5007 emission is an order of magnitude lower than that derived using H$\alpha$ emission.
This is likely a consequence of [OIII] emission arising from a smaller volume of gas compared to Balmer line emission and tracing only higher ionization state of the gas \citep[see][for a more detailed explanation]{carnianiIonisedOutflows242015}.

\begin{figure*}[h!]
    \centering
    \includegraphics[clip=true, trim=0.6cm 0.1cm 1.5cm 0.2cm,width=0.32\textwidth]{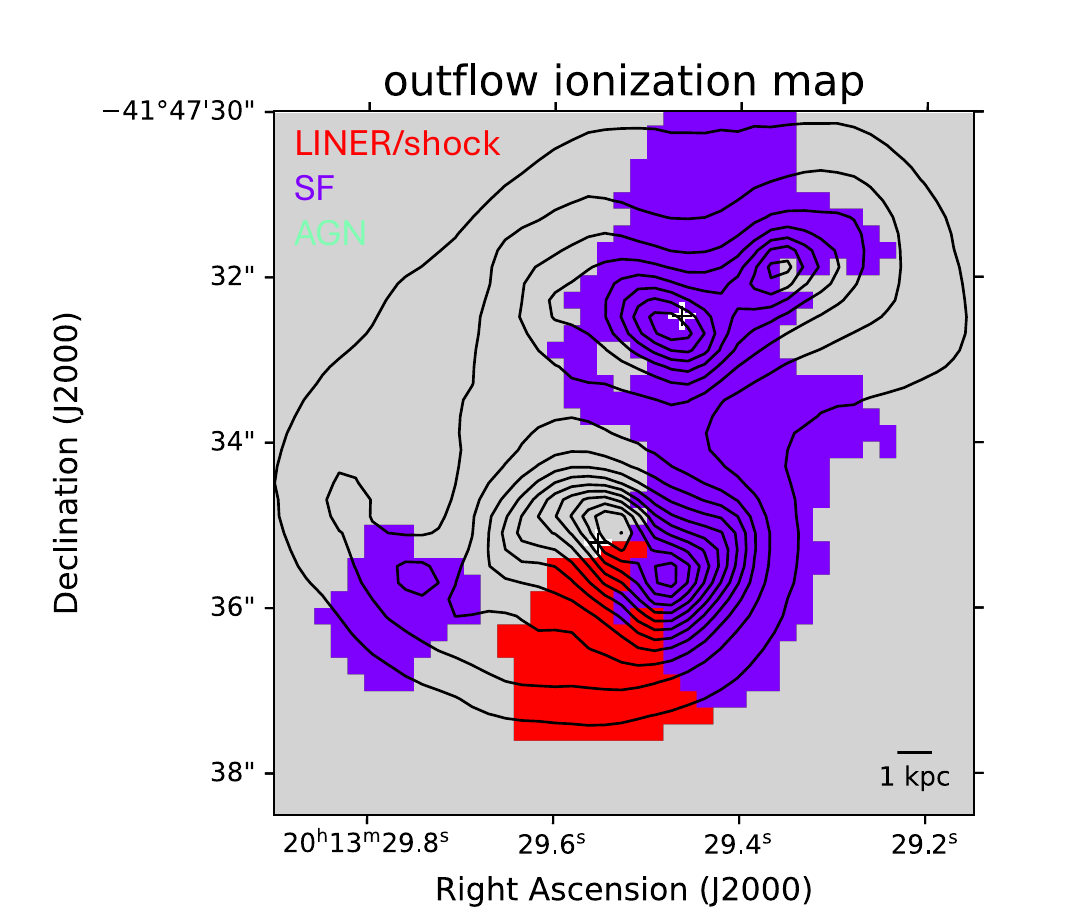}
    \includegraphics[clip=true, trim=0.1cm 0.3cm 0.1cm 0cm,width=0.33\textwidth]{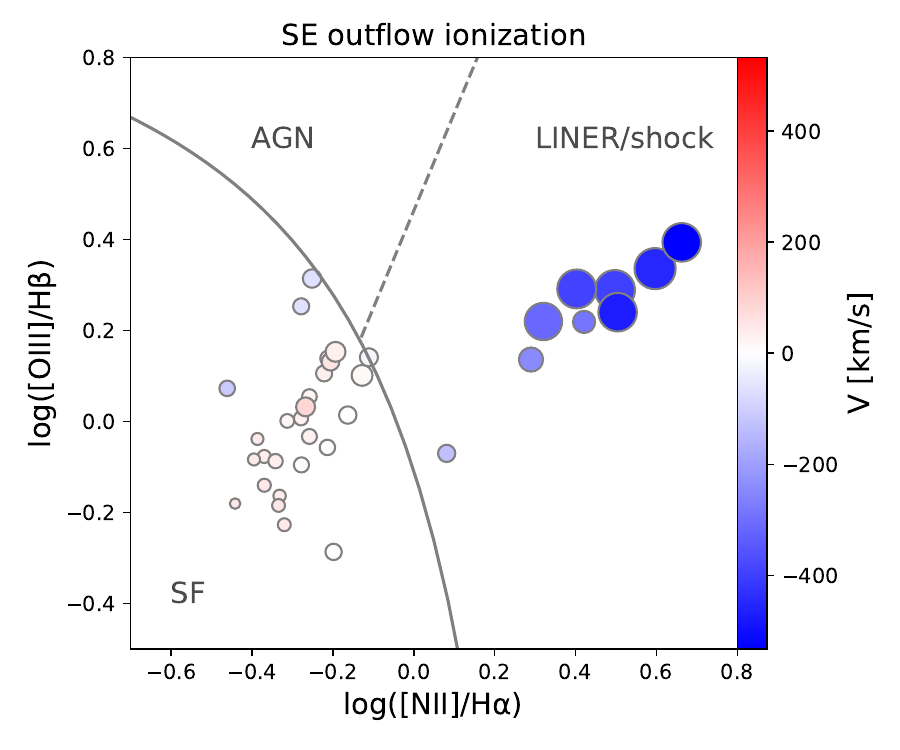}
    \includegraphics[clip=true, trim=0.1cm 0.3cm 0.1cm 0cm,width=0.33\textwidth]{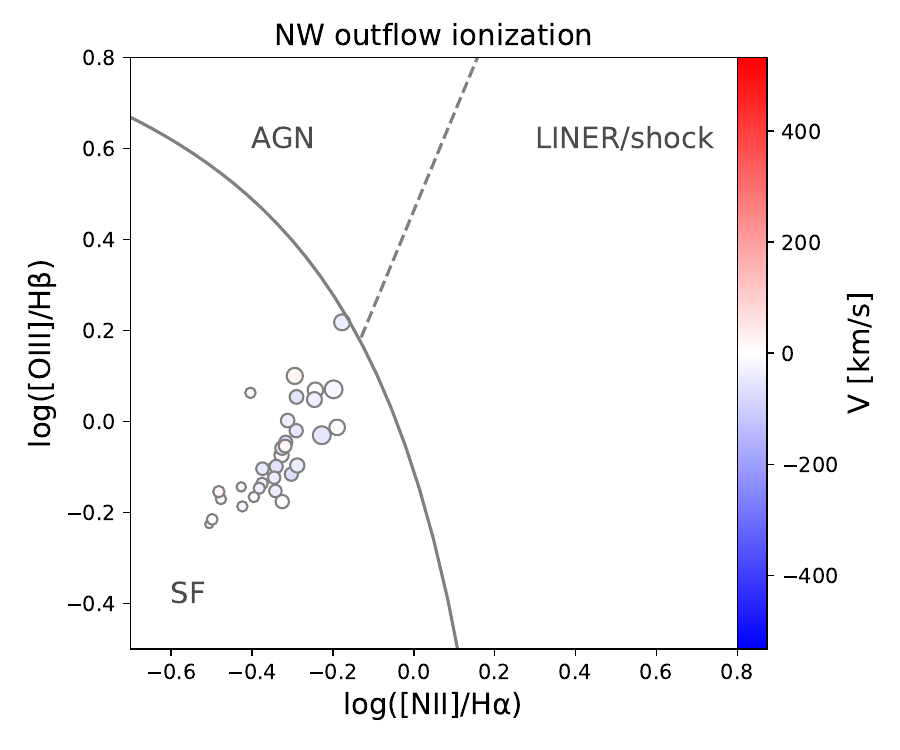}
    \caption{BPT classification of outflowing gas by Voronoi bins. In the maps, purple bins are dominated by SF/composite-like, red bins by shock and LINER-like, and mint bins by AGN-like (Seyfert) ionization. Contours correspond to the stellar continuum flux. In the BPT diagrams symbol color corresponds to central velocity of the fit component and symbol size (diameter) corresponds to velocity dispersion $\sigma$. The solid line shows the theoretical maximum for ionization due to star-formation from \citet{kewleyTheoreticalModelingStarburst2001} and the dashed line shows the demarcation between Seyfert (above the line) and LINER- or shock-like ionization (below the line) suggested by \cite{kauffmannHostGalaxiesActive2003}.}
    \label{fig:bpt_maps_vbin}
\end{figure*}

\subsection{Outflow ionization state and metallicity}
\label{sec:results:bpt_met}
We calculated the emission line ratios based on the line fluxes obtained from the fit to the MUSE data for each Voronoi bin.
From the line ratios, we were able to classify the ionization state of both the outflowing and quiescent components of the gas in each bin, based on the commonly used BPT diagram \citep{baldwinClassificationParametersEmissionline1981}.
For the quiescent component, the classification shows star formation is the dominant source of excitation throughout the system.
Figure~\ref{fig:bpt_maps_vbin} shows the resulting map, as well as the position in the diagram for each bin in which the relevant lines are detected.
The secondary component shows some signs of shock or LINER-like ionization in the SE, where the strongest outflows are present.
There is a clear bimodal distribution of the secondary components in the BPT diagram, with one cluster corresponding to lower-velocity and narrower secondary components situated in the star formation-like ionization region of the diagram and a second cluster of higher-velocity and broader secondary components deep in the shock- or LINER-like ionization region.
The metallicity was derived from the H$\alpha$, [NII]$\lambda6484$, and [SII]$\lambda\lambda6717,31$ emission lines according to the calibration in \citet{dopitaChemicalAbundancesHighredshift2016}. 
We used the quiescent line components to derive metallicities for the outflow in each bin, the shock- or LINER-like ionization of the outflowing gas means the strong line metallicity measures cannot reliably be applied there.
For the solar reference abundance $\mathrm{[O/H]_\odot}$ we adopted a value of 8.77, to be consistent with \citet{dopitaChemicalAbundancesHighredshift2016}.
{\bx The resulting metallicity maps are shown in Appendix~\ref{app:metallicity}.}

\subsection{Neutral atomic gas}
\begin{figure}
    \centering
    \includegraphics[clip=true, trim=0.4cm 0.1cm 0.1cm 0.1cm,width=0.4\textwidth]{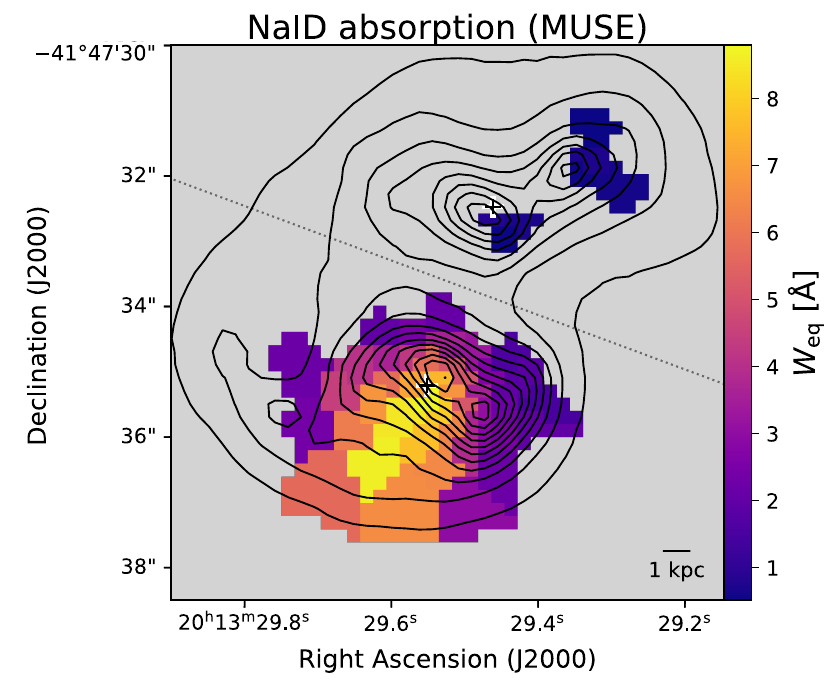}
    \caption{{\bx Equivalent width of the NaI doublet absorption features based on best-fit parameters. The black contours show the stellar continuum emission integrated over the spectral range.}}
    \label{fig:vorbin_neut_eqw_map}
\end{figure}

We used the best-fit Doppler parameter for the blueshifted NaI doublet nebular absorption feature to derive outflow velocities of the neutral atomic gas phase in each bin.
Panel (g) in Fig.~\ref{fig:all_phases_vorbin_vel_maps} shows the resulting velocity map, while Fig.~\ref{fig:vorbin_neut_eqw_map} shows a map of the EW of the absorption feature.
The blueshifted absorption overlaps with the blueshifted emission in CO(1-0) and H$\alpha$, but is somewhat more extended than the latter.
Summing contributions from all Voronoi bins associated with the SE nucleus and in which at least one NaI doublet component is detected with an $|A_\mathrm{G}|/N_\mathrm{G}>3$ yields the total mass of the SE neutral atomic outflow.
At $51\times10^7\msun$, the neutral outflow phase carries about five times the mass of the ionized phase traced by H$\alpha$ emission and is about 15\% as massive as blueshifted component of the cold molecular phase. We performed the same analysis for the bins associated with the NW nucleus and found that the corresponding outflow mass is consistent with zero given the associated uncertainties.

\section{Discussion}
\label{sec:discussion}

\subsection{Outflow velocity structure}
\label{sec:discussion:outflow_velocity_structure}
\begin{figure*}
    \centering
    \includegraphics[clip=true, trim=0.1cm 0.5cm 0.1cm 0.2cm, width=0.45\textwidth]{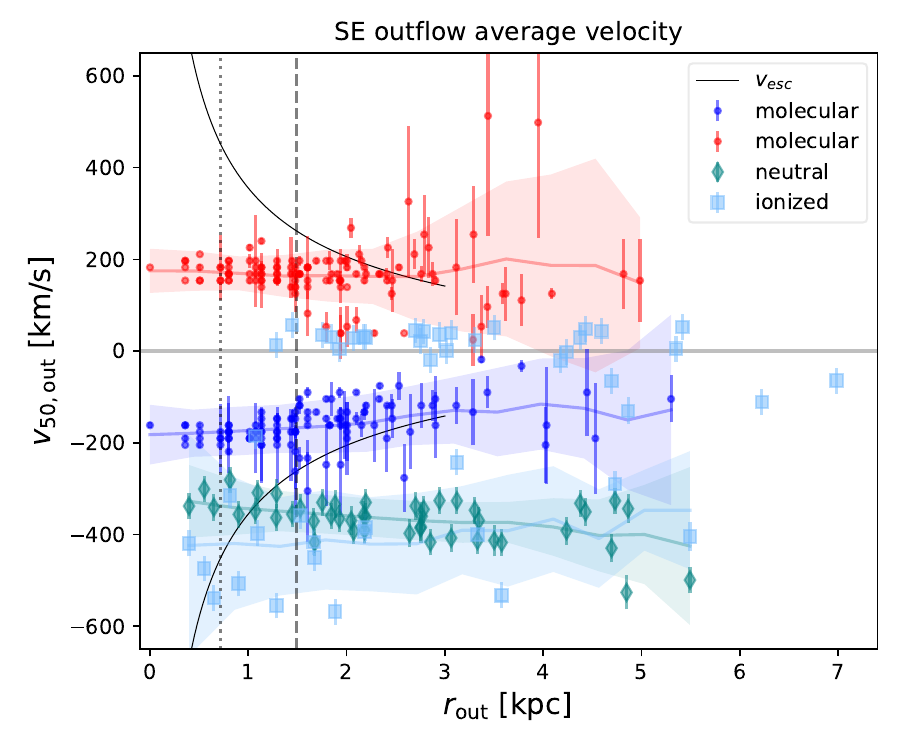}
    \includegraphics[clip=true, trim=0.1cm 0.5cm 0.1cm 0.2cm, width=0.45\textwidth]{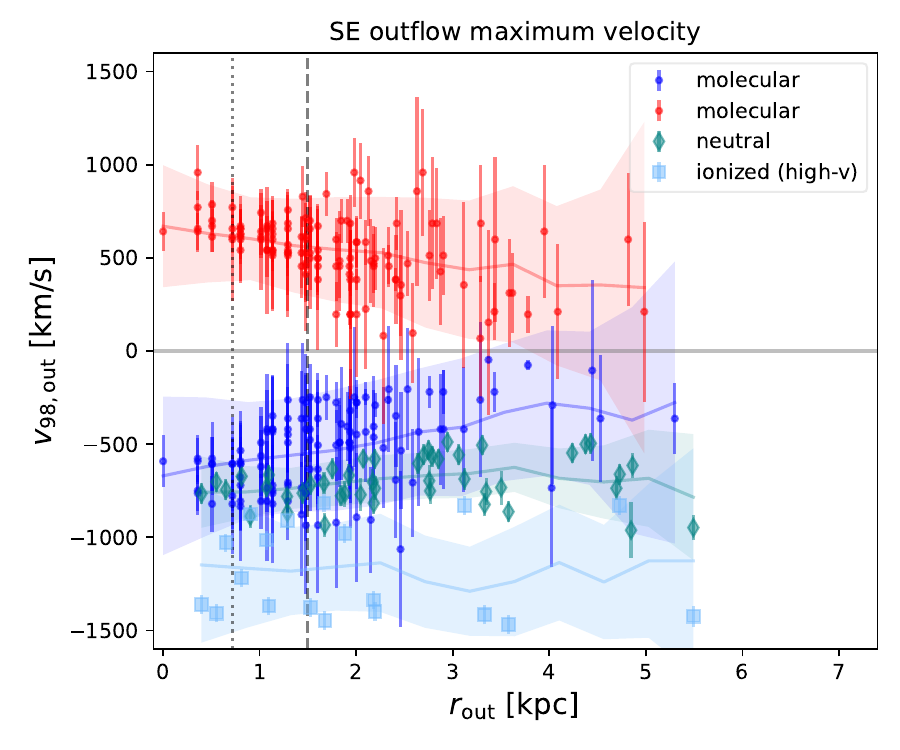}
    \caption{Average (left) and maximum (right) outflow velocities vs. distance from center for each Voronoi bin. Blue (red) dots show the blueshifted (redshifted) molecular outflow, teal diamonds show the neutral outflow, and light blue squares show the ionized outflow. The dashed (dotted) vertical line shows the adopted lower limit on $r_\mathrm{out}$ based on the ALMA (MUSE) angular resolution. We also show the running mean (with a uniform 2\,kpc kernel) velocity for each phase (only the high-v bins for the ionized gas) as a solid line and its estimated uncertainty as a shaded area in the corresponding colors. {\bx The solid black curves in the left panel indicate the estimated escape velocity based on the dynamical mass of the SE merger component from \citet{pernaPhysicsULIRGsMUSE2022}.}}
    \label{fig:rout_vout_bins}
\end{figure*}

Calculating the outflow velocities separately for each Voronoi bin allows us to investigate the structure of the outflow in different phases, in addition to computing global outflow properties.
We focus on the SE outflow for this analysis, where all three phases are clearly detected.
Figure~\ref{fig:rout_vout_bins} shows the average and maximum outflow velocities in each bin plotted against the distance from the SE nucleus for the molecular, neutral and ionized gas phases.
\begin{table*}[]
\small
\caption{Total outflow properties by components, gas phases, and tracers.}
    \centering
    \begin{tabular}{lccccccc}
        \hline
        \hline\\[-8pt]
         Outflow component & Tracer & $\langle v\rangle\,[\mathrm{\frac{km}{s}}]$& $v_\mathrm{max}\,[\mathrm{\frac{km}{s}}]$ & $\langle r_\mathrm{out}\rangle\,[\mathrm{kpc}]$ & $r_\mathrm{out,max}\,[\mathrm{kpc}]$ & $M_\mathrm{out}\,[10^7\,M_\odot]$ & $\dot{M}_\mathrm{out}\,[\frac{M_\odot}{\mathrm{yr}}]$\\[3pt]
         \hline
molecular blue SE & CO(1-0)     & $160$ & $310$ & 1.7 & 5.3 & $340\pm 100$ & $300\pm 90$ \\
molecular red SE & CO(1-0)      & $170$ & $510$ & 1.8 & 5.0 & $430\pm 130$ & $420\pm 130$ \\
molecular total SE & CO(1-0)    & $170$ & $510$ & 1.7 & 5.3 & $800\pm 200$ & $700\pm 200$ \vspace{0.2cm} \\ 
neutral SE & NaI                & $360$ & $500$ & $2.5$ & $5.5$ & $51\pm 12$ & $88\pm 14$ \vspace{0.2cm} \\ 
ionized SE (total) & H$\alpha$          & $190$ & $570$ & 2.9 & 7.0 & $9.2\pm 3$ & $11\pm 4.5$ \\
ionized SE (total) & $\mathrm{[OIII]}$  & $150$ & $540$ & 3.0 & 7.0 & $0.5\pm 0.31$ & $0.5\pm 0.30$ \\
ionized SE (high-v) & H$\alpha$ & 430 & $570$ & 2.1 & 5.5 & $2.2\pm 0.8$ & $10\pm 3.7$\\
ionized SE (high-v) & $\mathrm{[OIII]}$  & 400 & $540$ & 1.9 & 4.7 & $0.12\pm 0.07$ & $0.5\pm 0.30$\\
         \hline
    \end{tabular}
    \label{tab:multiphase_outflow_props}
\end{table*}

Outflow velocities in the ionized phase show a bimodal distribution, mirroring the bimodality in the ionization state of the Voronoi bins (see Sect.~\ref{sec:results:bpt_met}).
The cluster of bins with higher absolute outflow velocities ($\gtrsim 180$\,km/s) correlates with the cluster of bins displaying shock-like ionization (see Fig.~\ref{fig:bpt_maps_vbin}) and it forms the cone-like feature extending south-east of the SE nucleus, where we also observe the cold molecular and neutral outflow phases.
All this points to the possibility that it is only this high-velocity cluster that traces counterpart of the molecular outflow in the ionized gas, and that the lower velocity secondary components instead trace features that are not due to the outflow, but rather due to gravitationally induced motion in the merger with degenerate kinematics\footnote{We remind the reader that in contrast to the case of the molecular gas, for the ionized gas, we could not perform a proper decomposition into gravitational and non-gravitational features, due to the blending of H$\alpha$ and [NII] and to the low spectral resolution of the data.}.
Based on this, we compute ionized gas outflow properties for the high-velocity cluster bins separately and report them alongside the total for all bins (Table~\ref{tab:multiphase_outflow_props}).
In the blueshifted component of the SE outflow, where the cold molecular, neutral atomic and the ionized phases are all clearly present, we can compare their characteristic velocities. 
The cold molecular gas has the lowest characteristic velocity with 160\,km/s, followed by the neutral atomic phase with 360\,km/s, and the ionized phase with 430\,km/s (considering the high-velocity cluster traced by H$\alpha$).
Such differences between these outflow phases have been observed previously, for instance in the extensively studied multiphase outflow of M82 \citep[e.g.,][]{ShenLoM821995ApJ...445L..99S,ShopbellM821998ApJ...493..129S,WalterWeissM822002ApJ...580L..21W}. 
This observed velocity structure is consistent with theoretical predictions for the entrainment of cold gas by a hot outflow \citep{wardAGNdrivenOutflowsClumpy2024}, whereas cold gas formed by cooling in the outflow would be expected to have the same velocity as the hot phase from which it has formed \citep[see][for a review]{veilleuxGalacticWinds2005}.

The trends of characteristic outflow velocity with distance from the nucleus are most evident in the left panel of Fig.~\ref{fig:rout_vout_bins}, where we show the running mean for each phase calculated with a uniform kernel of 2\,kpc width.
There is no significant decrease of projected outflow velocities with distance from the nucleus out to the limit at which we detect outflowing gas ($\sim5.5\,$kpc) in any of the three phases.
Such a decrease might be expected since the gas will lose kinetic energy as it moves further up in the potential well of the galaxy.
In addition, the outflowing gas is likely to experience deceleration due to ram pressure, and potentially entrainment of additional material from the ISM.
The absence of a decreasing trend in velocity with distance thus suggests the presence of an ongoing acceleration mechanism acting on the outflowing gas out to kpc scales.
While the maximum outflow velocities ($v_\mathrm{98,out}$) may be said to show a mild decreasing trend with distance from the nucleus in the molecular gas (right panel of Fig.~\ref{fig:rout_vout_bins}), this measure is very sensitive to the signal-to-noise level which decreases with distance from the nucleus.
The highest velocity gas in the outflow corresponds to the faint tail of the emission line, which will be the first part to drop below the detection limit.
This will necessarily bias the maximum velocity measure to lower values in noisier Voronoi bins.
The 50th percentile velocity $v_\mathrm{50,out}$ is less affected by the potentially missed flux from high-velocity gas in the noisier bins and therefore should be a more robust measure of the characteristic outflow velocity.
{\bx 
The characteristic velocities of the neutral atomic gas phase we find here are consistent with previous results based on spectra extracted from the SE nucleus of IRAS20100-4156.
\citet{lampertiPhysicsULIRGsMUSE2022} and \citet{fluetschPropertiesMultiphaseOutflows2021} both find neutral outflow velocities of around 400\,\kms.
For the ionized phase, \citet{fluetschPropertiesMultiphaseOutflows2021} report a higher outflow velocity (900\,\kms) for the SE nucleus of IRAS20100-4156 than what we find here.
This discrepancy is likely a consequence of the definition they use for the characteristic outflow velocity based on \citet{rupkeOutflowsInfraredLuminousStarbursts2005}, in which half the FWHM of the outflow component is added to its central velocity.
Reported outflow velocities of the cold molecular phase in IRAS20100-4156 based on CO(1-0) measurements, vary substantially.
\citet{gowardhanDualRoleStarbursts2018} use a method based on a line decomposition with three Gaussian components which considers only gas at high velocities ($|v_\mathrm{out}| \gtrsim 400\,$km/s) as part of the outflow.
Unsurprisingly, this results in a significantly higher characteristic outflow velocity ($>900\,$km/s) than what we find here, but much lower flux attributed to the outflow (by a factor of 4-5 compared to this work).
\citet{fluetschColdMolecularOutflows2018} report a CO(1-0) based outflow velocity of $\sim450$\,\kms, using a double Gaussian decomposition of the emission line.
They adopt the definition for the characteristic outflow velocity from \citet{rupkeOutflowsInfraredLuminousStarbursts2005} mentioned above, which is likely the main driver of the higher outflow velocity they find.
}

{\bx 
The final fate of the outflowing molecular gas is somewhat uncertain.
Following \citet{lampertiPhysicsULIRGsMUSE2022}, we estimate the escape velocity for the SE merger component using Eq.~7 from \cite{arribasIonizedGasOutflows2014} with the dynamical mass within 3\,kpc of the nucleus based on the rotation curve modeling by \cite{pernaPhysicsULIRGsMUSE2022}.
Much of the molecular gas is found at velocities below the escape velocity (see Fig.~\ref{fig:rout_vout_bins}), but the velocity structure of the outflow suggests the gas is still being accelerated and may reach velocities sufficient for escape, as is the case at a radius of $\sim3$\,kpc.
Average outflow velocities of both the ionized and neutral phase exceed the estimated escape velocity at scales at which the outflow is resolved, indicating that the fraction of mass carried in these phases will escape the galaxy.
Following \cite{fluetschColdMolecularOutflows2018}, we use abundance matching \citep{mosterGalacticStarFormation2013} to estimate the dark matter halo mass and associated escape velocity, which exceeds 4000\,\kms, even at the largest extent at which the outflow is detected.
This indicates that the outflowing gas will remain in the gravitational potential of the halo.
}

\subsection{Outflow mass and mass rate}
\label{sec:outflow_mass_rates}
With outflow masses and velocities available for each gas phase, we can calculate their associated outflow mass rates, again separately for each Voronoi bin, as
\begin{equation}
\label{eq:mass_rate}
    \dot{M}_\mathrm{out} = \frac{M_\mathrm{out}v_\mathrm{out}}{r_\mathrm{out}},
\end{equation}
where $r_\mathrm{out}$ is the distance of each bin center to the SE nucleus.
For the cold molecular gas we impose a minimum $r_\mathrm{out}$ of 1.49\,kpc which corresponds to half the major axis (0.62'') of the synthesized ALMA beam in the CO(1-0) observations.
Similarly, for the outflow phases based on MUSE data (H$\alpha$, [OIII], and NaI), we enforce a minimum $r_\mathrm{out,ion}$ of 0.72\,kpc, equivalent to half the FWHM of the MUSE PSF for these observations.
Summing over the contributions from all bins, we obtain the total outflow mass rates without assuming a specific geometry of the outflow.
The results are listed in Table~\ref{tab:multiphase_outflow_props} alongside typical and maximum values of outflow extent and velocity for the different components and gas phases.
Notably, when considering only the high-velocity cluster of bins for the ionized phase of the outflow, the resulting outflow mass rate does not change significantly compared to the sum of all bins ($10\,\msun$/yr compared to $11\,\msun$/yr). 
The cold molecular phase contributes by far the most to the total mass outflow rate with 700\,$\msun$/yr.
This mass rate corresponds to a mass loading factor $\eta = \dot{M}_\mathrm{out,H_2}/\mathrm{SFR}$ of 3.5, which is within what is observed for local "pure" starburst ULIRGs \citep{ciconeMassiveMolecularOutflows2014}. 
\citet{gowardhanDualRoleStarbursts2018} find almost exactly the same mass outflow rate in their resolved study of CO(1-0) emission in IRAS20100-4156, despite their very different approach in identifying the outflow.
The higher outflow velocity and lower flux attributed to the outflow in their work act in opposite directions when calculating mass outflow rates (see Eq.~\ref{eq:mass_rate}), explaining the similar results between the two approaches.
Despite the similar mass outflow rates, our approach nonetheless paints a very different picture of the outflow as a whole, with over 40\% of the total cold molecular gas mass in the system being affected by feedback and potentially removed from the galaxy.

{\bx
\citet{fluetschPropertiesMultiphaseOutflows2021} report mass rates for the neutral atomic phase of the outflow in IRAS20100-4156 very similar to what we find here
However, the mass rates they find for the ionized phase based on H$\alpha$ emission exceed our results by about a factor of 10.
Their results are based on a single integrated spectrum extracted from the MUSE cube, with an inferred electron density of 470\,cm$^{-3}$ for the outflowing gas.
In contrast, we find higher electron densities in some of the Voronoi bins with the highest ionized outflow velocities (compare Fig.~\ref{fig:electron_density_maps_vbin} and Fig.~\ref{fig:all_phases_vorbin_vel_maps}).
The contribution of these bins to the overall outflow mass and mass rate is thus decreased, which could explain the observed difference in ionized outflow mass rate.
}

\subsection{Outflow energetics}
\label{sec:discussion:outflow_energetics}
While the relatively slow and very massive molecular outflow we detect produces mass rates similar to those found in previous studies of IRAS20100-4156, its momentum rate ($\dot{p}_\mathrm{out}=\dot{M}_\mathrm{out}v_\mathrm{out}$) and kinetic power ($\dot{E}_\mathrm{kin,out}=\dot{M}_\mathrm{out}v_\mathrm{out}^2/2$) will necessarily be significantly lower in comparison, due to these quantities depending on the outflow velocity quadratically and cubically.
We find a momentum rate of $8.1\times10^{35}\,\mathrm{dyn}$ and a kinetic power of $7.8\times10^{42}\,\mathrm{erg/s}$ for the cold molecular phase of the outflow.
\citet{gowardhanDualRoleStarbursts2018} report a momentum rate about five times larger than what we find here and the difference in kinetic power is a factor of $\sim20$, as expected.
The total kinetic power of the outflow is $12\times10^{42}\,\mathrm{erg/s}$ after adding contributions form the neutral ($3.8\times10^{42}\,\mathrm{erg/s}$) and ionized ($0.57\times10^{42}\,\mathrm{erg/s}$, high-v only) phases.
If we assume the theoretical prediction for the mechanical energy injection from supernovae and stellar winds of $7\times10^{41} (\mathrm{SFR}/[\msun \mathrm{yr^{-1}}])\,$erg/s from \cite{veilleuxGalacticWinds2005}, this would require a coupling efficiency of $\sim9\%$ for a starburst driven outflow, which is close to theoretical predictions of the order of 5\% \citep{walchEnergyMomentumInput2015}.

The total momentum rate of the outflow across all phases is $10\times10^{35}\,\mathrm{dyn}$, with the neutral atomic phase contributing $2.0\times10^{35}\,\mathrm{dyn}$ and the ionized phase $0.3\times10^{35}\,\mathrm{dyn}$.
With a theoretically predicted momentum injection rate of $5\times10^{33} (\mathrm{SFR}/[\msun \mathrm{yr^{-1}}])\,$dyn, this would correspond to a net momentum transfer on the order of 100\% if the outflow were driven purely by mechanical stellar feedback.
However, there may be contributions from radiation pressure on dust in the outflowing gas \citep{thompsonDynamicsDustyRadiationpressuredriven2015}.
Indeed, for IRAS20100-4156, the radiation pressure, $\tau L_\mathrm{bol}/c$, could be of the same order of magnitude as the pressure due to mechanical feedback, assuming an optically thick outflow ($\tau \approx 1$) and $L_\mathrm{bol} \approx L_\mathrm{IR}$.
In ULIRGs, radiation pressure on dusty clouds can accelerate them to several hundreds of km/s \citep{zhangDustyCloudAcceleration2018}.
Since the cold molecular phase dominates the mass rate and energetics, our choice of CO-to-H$_2$ conversion factor $\alpha_\mathrm{CO}$ will introduce a significant additional systematic uncertainty.
For cold molecular gas in outflows, $\alpha_\mathrm{CO}$ is highly uncertain and may be lower than the value we adopt by a factor of a few due to an increased velocity dispersion of the outflowing gas \citep{pereira-santaellaCOtoH2ConversionFactor2024}.
This effect may, however, be counteracted by the dissociation of CO in the outflowing gas \citep{ciconeALMA3132018}, the extent of which is also uncertain.
Due to this, the comparisons and derived efficiencies in this section should be considered as order-of-magnitude estimates.

\subsection{Outflow origin}
\label{sec:discussion:outflow_origin}
The stellar and ionized gas velocity maps (Figs.~\ref{fig:vorbin_vel_maps_stars} and \ref{fig:all_phases_vorbin_vel_maps}) suggest a relatively intact disk around the SE nucleus of the merger.
The molecular outflow we identify in the SE shows a morphology consistent with a bi-cone originating in the SE nucleus, and oriented along the minor axis of the disk.
\citet{gowardhanDualRoleStarbursts2018} find the same orientation of the outflow in their analysis of the CO(1-0) emission, but they conclude that the direction of the outflow is aligned with the galaxy-wide velocity gradient based on the 1st moment map of the ALMA cube.
This highlights the need for a more sophisticated decomposition of outflow and quiescent gas components in systems with powerful outflows, as moment maps can be significantly influenced by massive outflows, especially at intermediate spatial resolution.
The blueshifted outflowing components observed in the ionized gas tracers and NaI absorption match the blueshifted side of the cold molecular outflow in orientation and extent.
This suggests a multiphase outflow being driven by feedback in the SE nucleus.
In the NW nucleus, emission from non-quiescent gas is far weaker and the associated velocities are much lower than in the SE nucleus, in both the molecular and ionized phases.
The same is true for absorption from the neutral atomic phase.
In addition, none of these tracers show a morphology or kinematics clearly indicative of an outflow in the NW.
This is consistent with the findings of \citet{gowardhanDualRoleStarbursts2018}, who detect outflowing molecular gas only in the SE nucleus using CO(1-0) emission, and \citet{lampertiPhysicsULIRGsMUSE2022}, who do not detect a molecular outflow in the NW using CO(2-1) emission but derive an upper limit on the molecular outflow mass rate there.
The BPT classification shows that the dominant ionization mechanism for the quiescent emission line components throughout the galaxy is star formation.
For the outflowing components in Voronoi bins where the ionized gas phase of the outflow is clearly detected, the BPT classification instead suggests shock or LINER-like ionization.
Since only the line ratios in the outflow are inconsistent with star formation as a primary source of ionization, shock ionization seems the more likely explanation, as otherwise LINER-like ionization would also be expected in the quiescent gas in the same area.
This result is somewhat in contrast to previous reports of LINER-like ionization in the quiescent gas around the SE nucleus \citep[e.g.,][]{gowardhanDualRoleStarbursts2018, pernaPhysicsULIRGsMUSE2021, lampertiPhysicsULIRGsMUSE2022}, which could be due to the amount of flux in the outflow at low velocities skewing the line ratios in integrated spectra when using fixed velocity cuts for the outflow decomposition.
It is still possible that IRAS20100-4156 hosts a heavily obscured AGN, which could explain the ambiguities in its classification \citep{franceschiniXMMNewtonHardXray2003, SpoonMIRgalaxyClassification2007, lampertiPhysicsULIRGsMUSE2022}; however, the lack of evidence to support AGN activity directly affecting the outflowing gas in combination with the outflow mass rate and energetics being consistent with feedback from starburst activity alone suggests there is no significant AGN contribution to the driving of the SE outflow in IRAS20100-4156.

\subsection{Molecular gas depletion time}
\label{sec:discussion:depletion_time}
We can use the outflow mass rates to estimate the molecular gas depletion timescale $\tau_\mathrm{dep,mol}=M_\mathrm{mol}/\dot{M}_\mathrm{out,mol}$, which is the time it would take for the outflow to remove all molecular gas from the galaxy at the current rate.
We find a depletion timescale of 16\,Myr, which lies within the estimated lifetime of starburst activity in ULIRGs \citep{farrahStarburstAGNActivity2003, EfstathiouULIRGmodelling2022}.
This implies that the starburst driven outflow could plausibly quench the galaxy over the course of its existence by driving the observed outflow.
However, we cannot definitively conclude that the outflow will quench the galaxy without two major assumptions.
First, starburst activity would need to consistently drive an outflow of the proportions we observe here over the lifetime of the ULIRG.
Second, the outflow would need to effectively remove gas from the galaxy for the purpose of star formation, which given the low velocity of some of the outflowing molecular gas, may not be true for the total outflowing gas mass.

\subsection{Dust}
\label{sec:discussion:dust}
\begin{figure}[]
    \centering
    \includegraphics[clip=true, trim=0.25cm 0.1cm 0.9cm 0.1cm,width=0.5\textwidth]{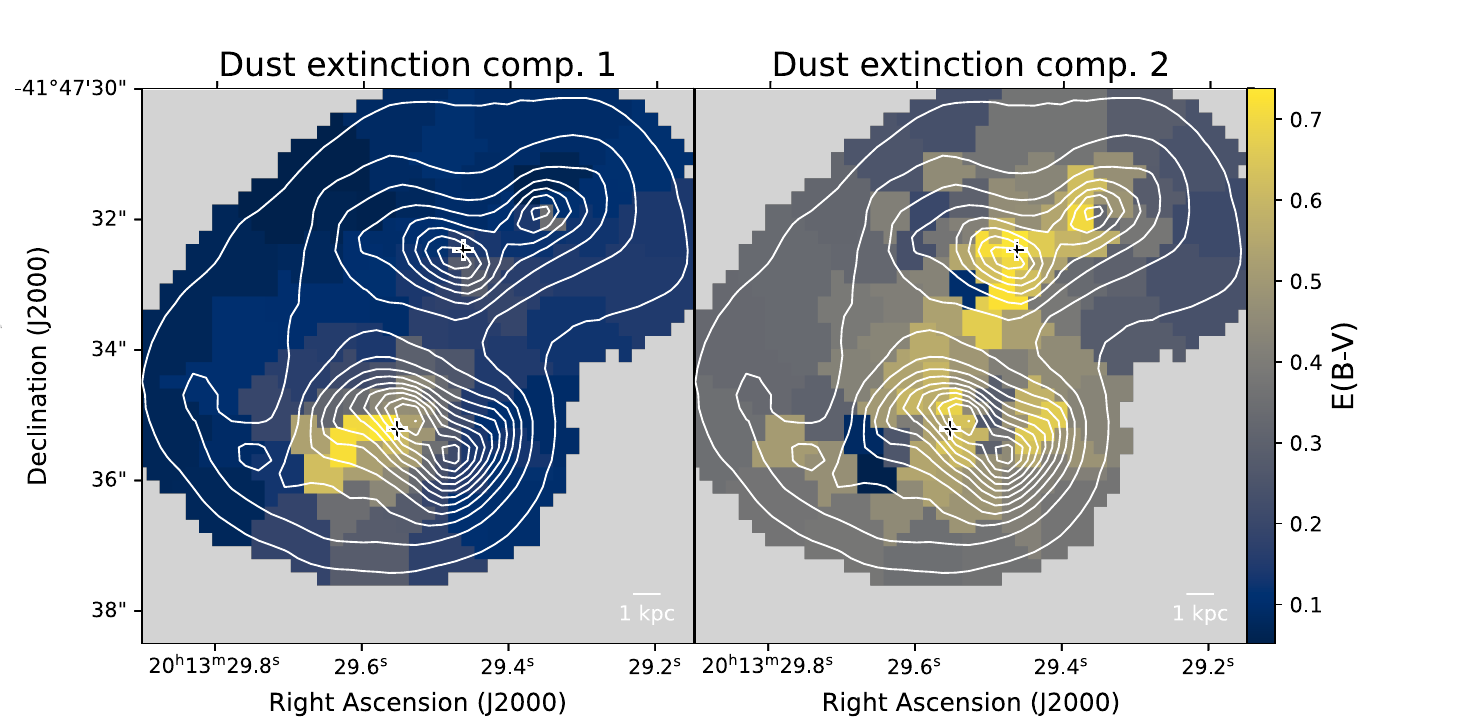}
    \caption{Map of the fitted dust extinction components, where component 1 (left) affects both nebular and stellar emission and component 2 (right) affects nebular emission only. The black contours show the stellar continuum emission integrated over the spectral range.}
    \label{fig:vorbin_dust_maps}
\end{figure}
Dust extinction is accounted for in the fit to the MUSE data with a two component model based on \citet[see our Appendix~\ref{app:gandalf} for details]{calzettiDustContentOpacity2000}.
The first of these components accounts for diffuse dust between the galaxy and the observer attenuating the full spectrum (stellar continuum and emission lines) in the rest-frame optical, while the second accounts for dust mixed with the emitting gas, attenuating only the nebular emission.
Extinction measured by second component of the dust model generally correlates well with the continuum emission (see Fig.~\ref{fig:vorbin_dust_maps}).
The first component, in stark contrast, shows high extinction exclusively in those regions where the blueshifted SE outflow is detected in the gas tracers (see Fig.~\ref{fig:vorbin_dust_maps}).
It is therefore likely that this model component traces primarily dust in the outflow in this galaxy.

\begin{figure}[]
    \centering
    \includegraphics[clip=true, trim=0.6cm 0.1cm 0.3cm 0.1cm,width=0.4\textwidth]{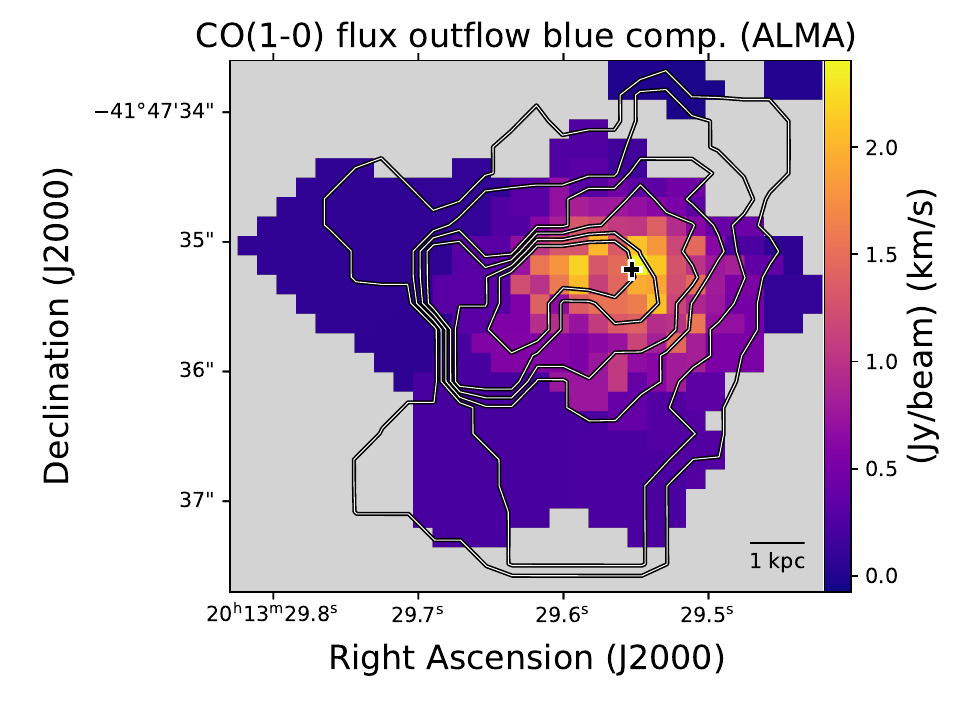}
    \includegraphics[clip=true, trim=0cm 0cm 0cm 0cm,width=0.38\textwidth]{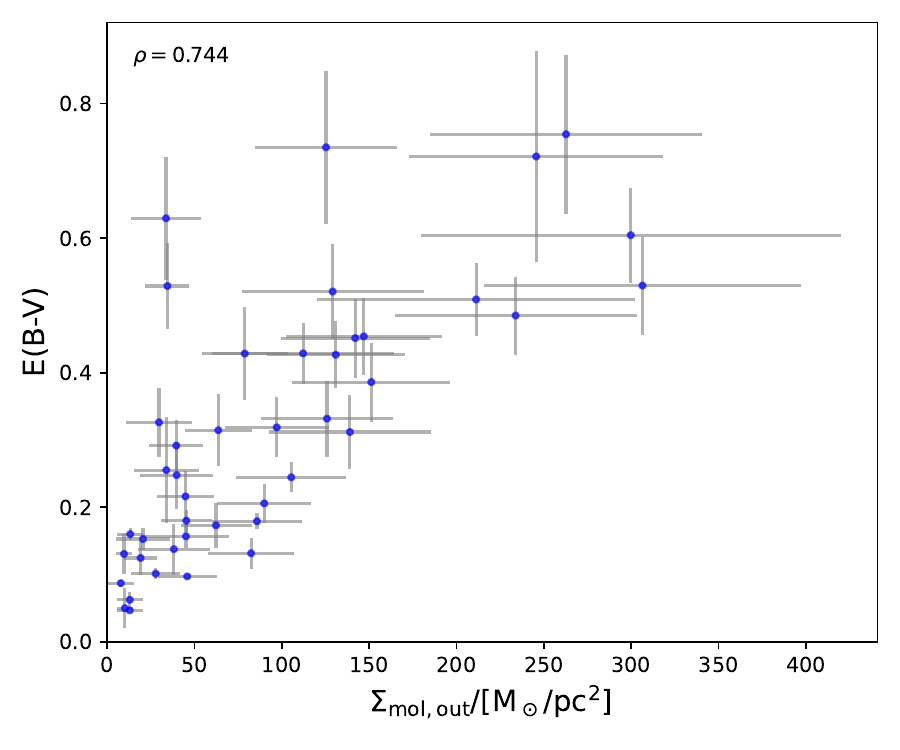}
    \caption{Comparison between blueshifted CO emission and diffuse dust extinction. Top: zoomed in view showing the flux in the blueshifted outflowing molecular gas component with contours indicating the dust extinction E(B-V) due to the diffuse dust component fitted to the MUSE data in the \texttt{GandALF} model (cf. Fig.~\ref{fig:vorbin_dust_maps}). Bottom: Surface density of blueshifted outflowing molecular gas vs. dust extinction E(B-V) in MUSE Voronoi bins.}
    \label{fig:co10_blue_nonrot_flux_dust}
\end{figure}

We test this scenario by comparing the surface density of outflowing cold molecular gas derived from ALMA CO(1-0) observations to the dust extinction E(B-V) from the fit to the MUSE data.
If the extinction component indeed arises from dust mixed with the outflowing cold molecular gas, and if the dust-to-gas mass ratio is approximately constant across the outflow, then we expect a correspondence between CO intensity, which scales with cold molecular gas surface density, and the extinction E(B-V), which scales with the dust surface density. 
The top panel of Fig.~\ref{fig:co10_blue_nonrot_flux_dust} shows the intensity of CO emission from outflowing gas overlayed with contours corresponding to dust extinction from the diffuse dust component, which match the CO emission quite well.
The bottom panel of Fig.~\ref{fig:co10_blue_nonrot_flux_dust} shows the markedly positive correlation between the surface density of molecular gas and E(B-V).
Dust extinction from outflowing gas has previously been linked to the neutral gas phase, though more commonly to atomic gas column densities \citep{rupkeMULTIPHASESTRUCTUREPOWER2013}.
Indeed, there is a correlation between the EW of NaID absorption and diffuse dust extinction in IRAS20100-4156 (compare Figs.~\ref{fig:vorbin_neut_eqw_map} and \ref{fig:HI_dust_corr}, and see Appendix~\ref{app:dust}.
However, we find the correlation with outflowing gas surface density to be much stronger for the cold molecular phase than for the neutral atomic phase (see Appendix~\ref{app:dust}), suggesting the majority of the dust is carried by the cold molecular gas.
Based on this, we assume that diffuse dust extinction is primarily due to dust mixed with the outflowing cold molecular gas.
Under this assumption, we can estimate the mass of dust in the outflow using a metallicity dependent dust-to-gas ratio from \citet{draineInfraredEmissionInterstellar2007}:
\begin{equation}
    M_\mathrm{dust,out} = 0.01\times 10^{[\mathrm{O/H}]-[\mathrm{O/H}]_\odot} M_\mathrm{mol,out},
\end{equation}
where $[\mathrm{O/H}]$ is the gas-phase metallicity in the outflow, $[\mathrm{O/H}]_\odot$ is the Milky Way value, which we assume to be 8.77, and $M_\mathrm{mol,out}$ is the total mass of the cold molecular outflow.
This results in an outflowing dust mass of $3.5\times 10^7\,M_\odot$ detected in the blueshifted part of the molecular outflow or a total outflowing dust mass of $8.3\times 10^7\,M_\odot$ if we assume that the redshifted part of the outflow carries the same fraction of dust.
Similar dust masses ($10^6$-$10^7\msun$) have been inferred from observations of extra-planar dust around starburst galaxies on scales of 2-10\,kpc \citep{Hodges-KluckExtraplanarDust2016}.
The presence of dust in the outflowing gas is further evidence that radiation pressure may play a role as a driving mechanism for the outflow in IRAS20100-4156.

\section{Summary and conclusion}
\label{sec:conclusion}
In this study, we investigate the massive multiphase outflow in the galaxy merger IRAS20100-4156.
We used the stellar kinematics derived from a full-spectrum fit to rest-frame optical MUSE observations to define a velocity field dominated by gravitational motions and identify a matching component in the complex CO(1-0) emission line profiles observed with ALMA.
We subtracted this component tracing quiescent gas and analyzed the remaining emission assuming it arises from outflowing gas.
This procedure revealed a cold molecular outflow in the SE merger component of IRAS20100-4156 with the following characteristics:
\begin{itemize}
    \item very massive, with a mass of $8\times10^{9}\,M_\odot$, and accounts for about 40\% of the total cold molecular gas mass in the system.
    \item  an average outflow velocity of 170\,km/s, which is relatively slow compared to the neutral and ionized phases.
    \item  a bi-cone-like morphology centered on the SE nucleus and extending out to 5.5\,kpc with about equal mass and mass rates in the blue and redshifted cones.
    \item an estimated $8.3\times 10^7\,M_\odot$ of dust, identified by its extinction of the optical spectra.
\end{itemize}
There is a significant contribution of flux to the molecular outflow from gas at relatively low velocities, but with kinematics that are nonetheless inconsistent with quiescent gravitationally induced motion.
This component would be missed by outflow identification techniques that rely on fixed velocity cuts to decompose line emission into outflowing and quiescent components.
Co-spatially with the blueshifted side of the cold-molecular gas outflow, we detected outflow in the ionized gas (through H$\alpha$ and [OIII]$\lambda$5007 emission) and neutral atomic gas (through NaI doublet absorption).
Both of these phases are less massive than the cold molecular phase (containing 3\% and 15\% of its mass respectively), but exhibit faster average outflow velocities.
The emission line ratios in the outflowing gas is consistent with shock ionization, while the quiescent gas shows star formation-like ionization.
We used the calculated outflow masses and velocities to compute the momentum rate and kinetic power of the outflow.
The derived outflow energetics are dominated by the cold molecular phase,  consistently with theoretical predictions for an outflow driven by star formation through mechanical feedback from supernovae and stellar winds in addition to radiation pressure.
This leads us to the conclusion that the outflow is most likely driven by the starburst activity in the SE nucleus. Furthermore, if there is a heavily obscured AGN present in the source, its contribution to driving the observed outflow is not likely to be significant.
Finally, the outflow does not appear to slow down significantly with distance from its origin out to the limit of the extent at which it is detected.
This suggests there is an acceleration mechanism acting on the outflowing gas on kiloparsec scales.

\begin{acknowledgements}
    We would like to thank the anonymous referee for their insightful comments.
    This paper makes use of the following ALMA data: ADS/JAO.ALMA\#2013.1.00659.S, ADS/JAO.ALMA\#2018.1.00888.S. ALMA is a partnership of ESO (representing its member states), NSF (USA) and NINS (Japan), together with NRC (Canada), NSTC and ASIAA (Taiwan), and KASI (Republic of Korea), in cooperation with the Republic of Chile. The Joint ALMA Observatory is operated by ESO, AUI/NRAO and NAOJ.
    Parts of this work are based on observations made with ESO Telescopes at the La Silla Paranal Observatory under programme ID 0103.B-0391.
    CC acknowledges funding from the European Union's Horizon Europe research and innovation programme under grant agreement No. 101188037 (AtLAST2).
    CC, PS and CV  acknowledge a financial contribution from the Bando Ricerca Fondamentale INAF 2022 Large Grant, `Dual and binary supermassive black holes in the multimessenger era: from galaxy mergers to gravitational waves' and from the INAF Bando Ricerca Fondamentale INAF 2024 Large Grant: `The Quest for dual and binary massive black holes in the gravitational wave era'.
\end{acknowledgements}

\bibliography{Project2_sources}

\appendix

\section{CO(1-0) and CO(3-2) moment maps}
\label{app:CO_moment_maps}
\begin{figure}[!h]
    \centering
    \includegraphics[clip=true, trim=1.1cm 0.1cm 1.1cm 0.1cm,width=0.48\textwidth]{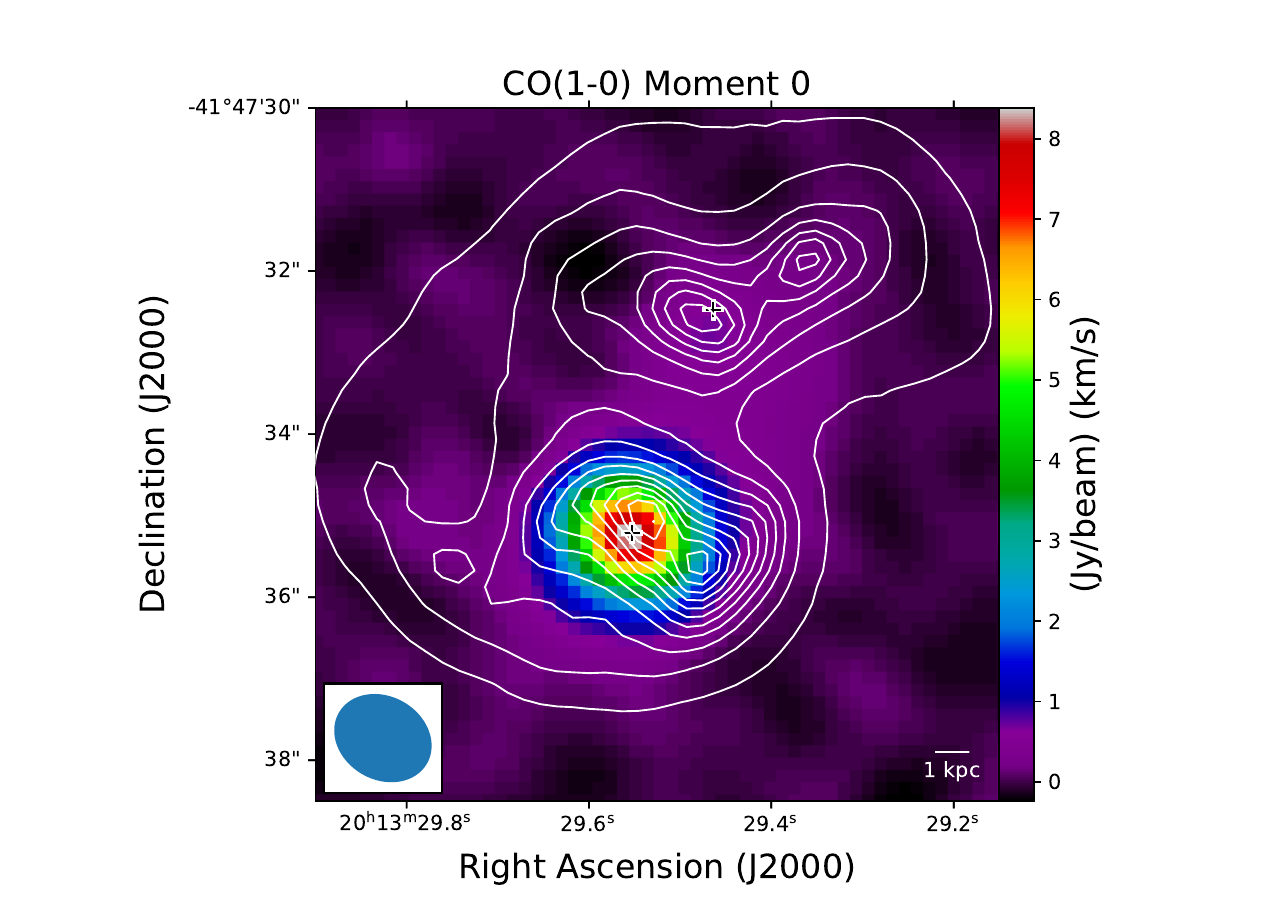}
    \includegraphics[clip=true, trim=1.1cm 0.1cm 1.1cm 0.1cm,width=0.48\textwidth]{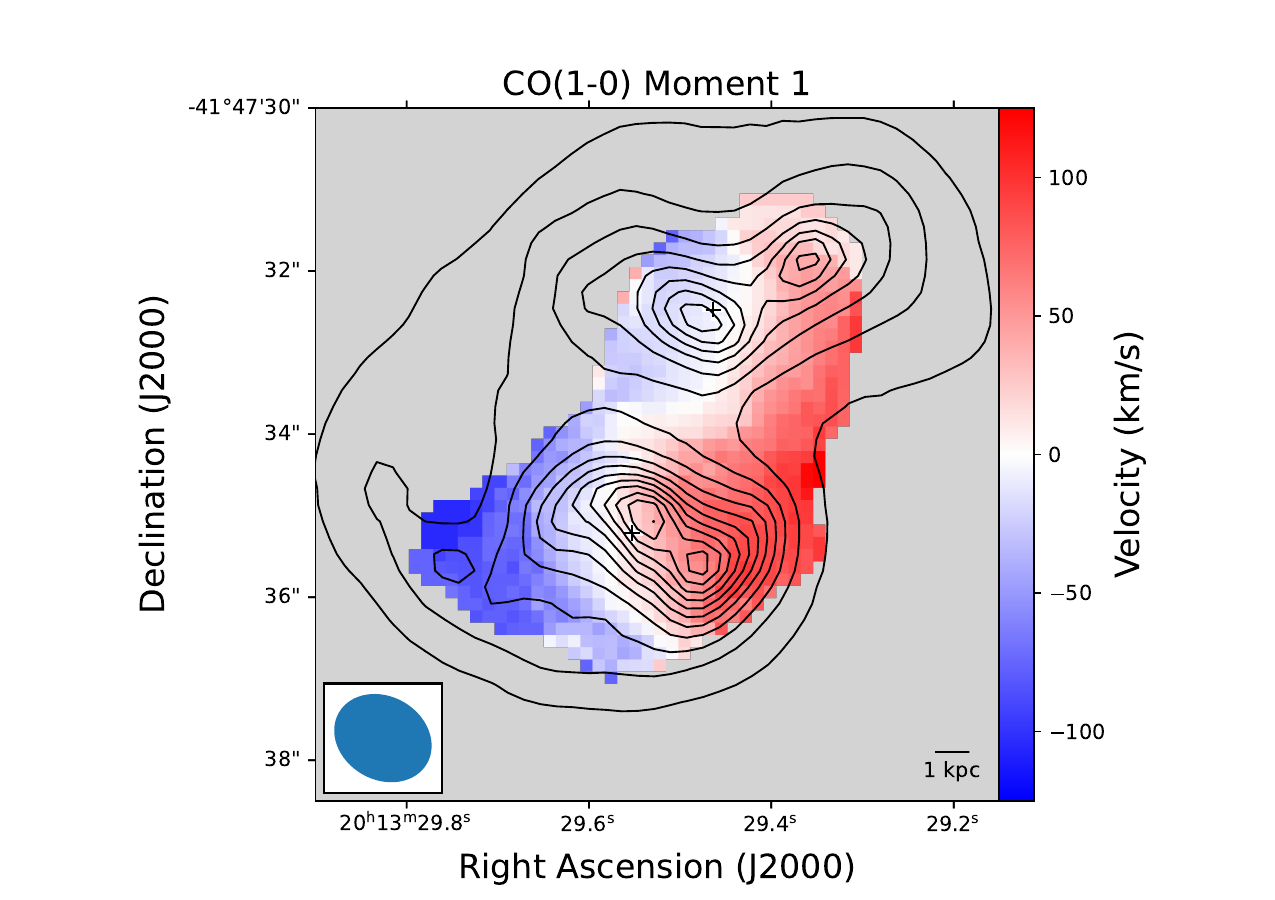}
    \includegraphics[clip=true, trim=1.1cm 0.1cm 1.1cm 0.1cm,width=0.48\textwidth]{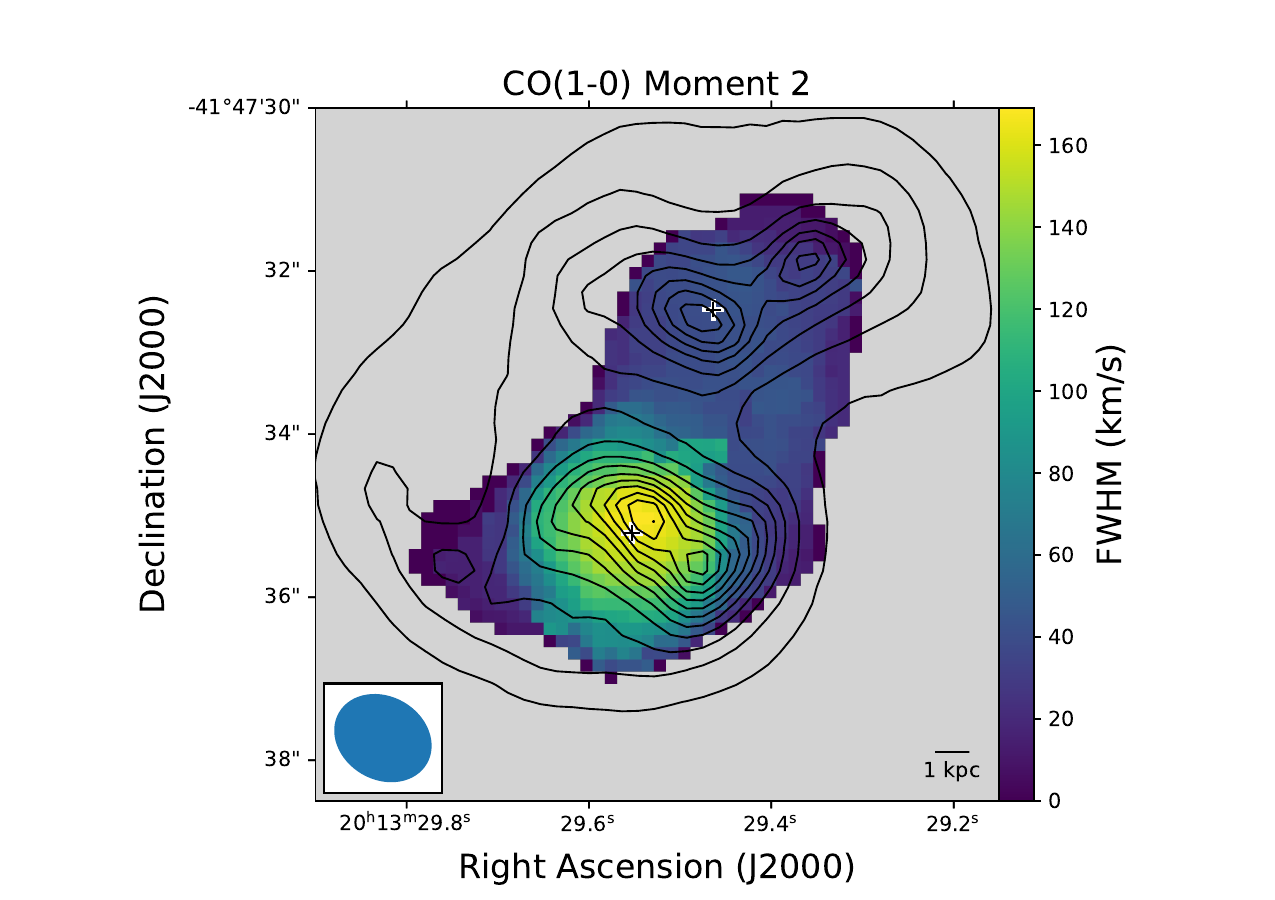}
    \caption{Moment maps of ALMA CO(1-0) data. The white (black in the second and third panels) contours show the stellar continuum emission observed with MUSE. The black and white markers show the positions of the CO(3-2) emission peaks in the SE and NW nuclei.}
    \label{fig:co_moment_maps}
\end{figure}

\section{MUSE cube fitting}
\label{app:gandalf}
\begin{figure}
    \centering
    \includegraphics[width=0.48\textwidth]{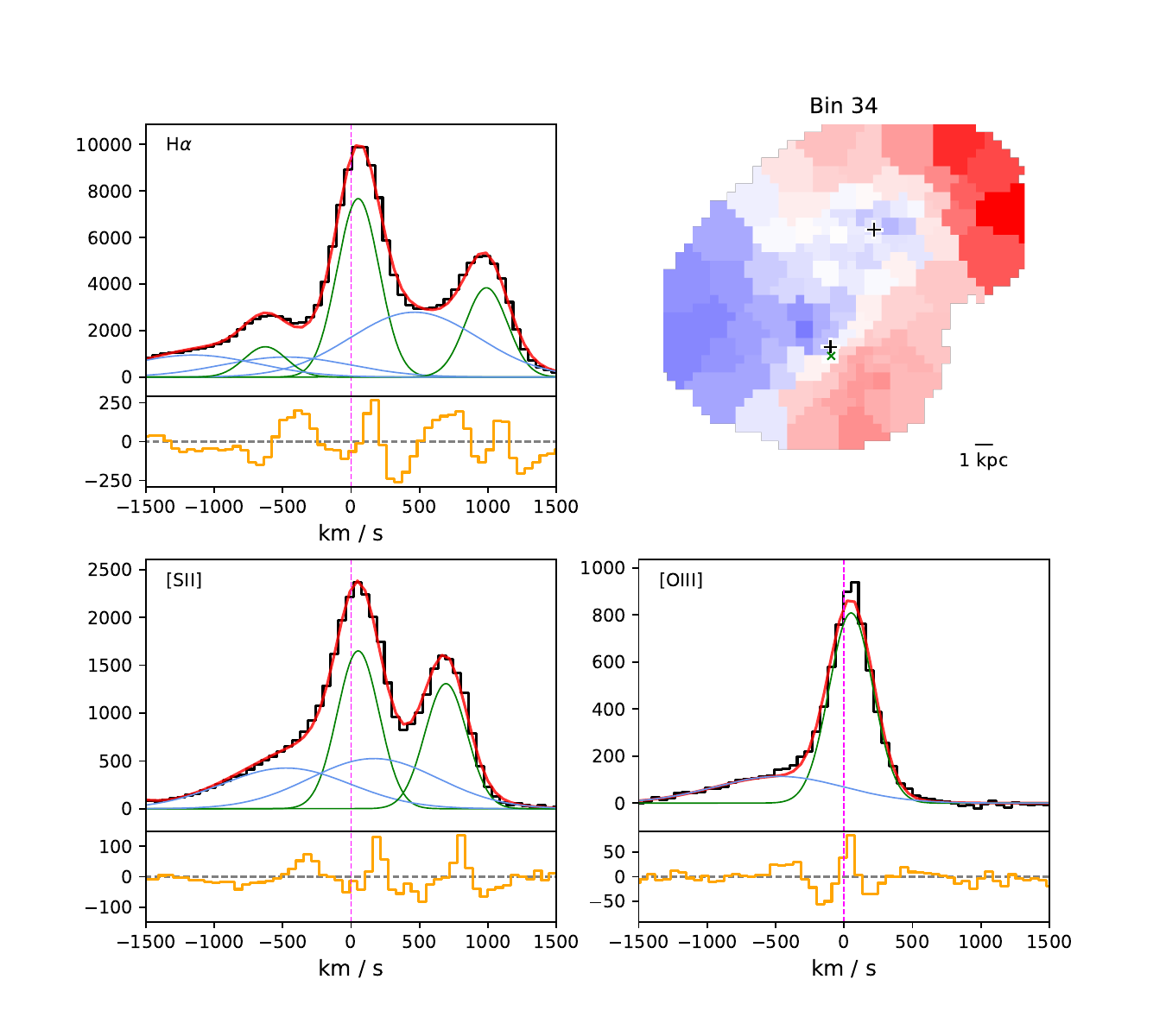}
    \includegraphics[width=0.48\textwidth]{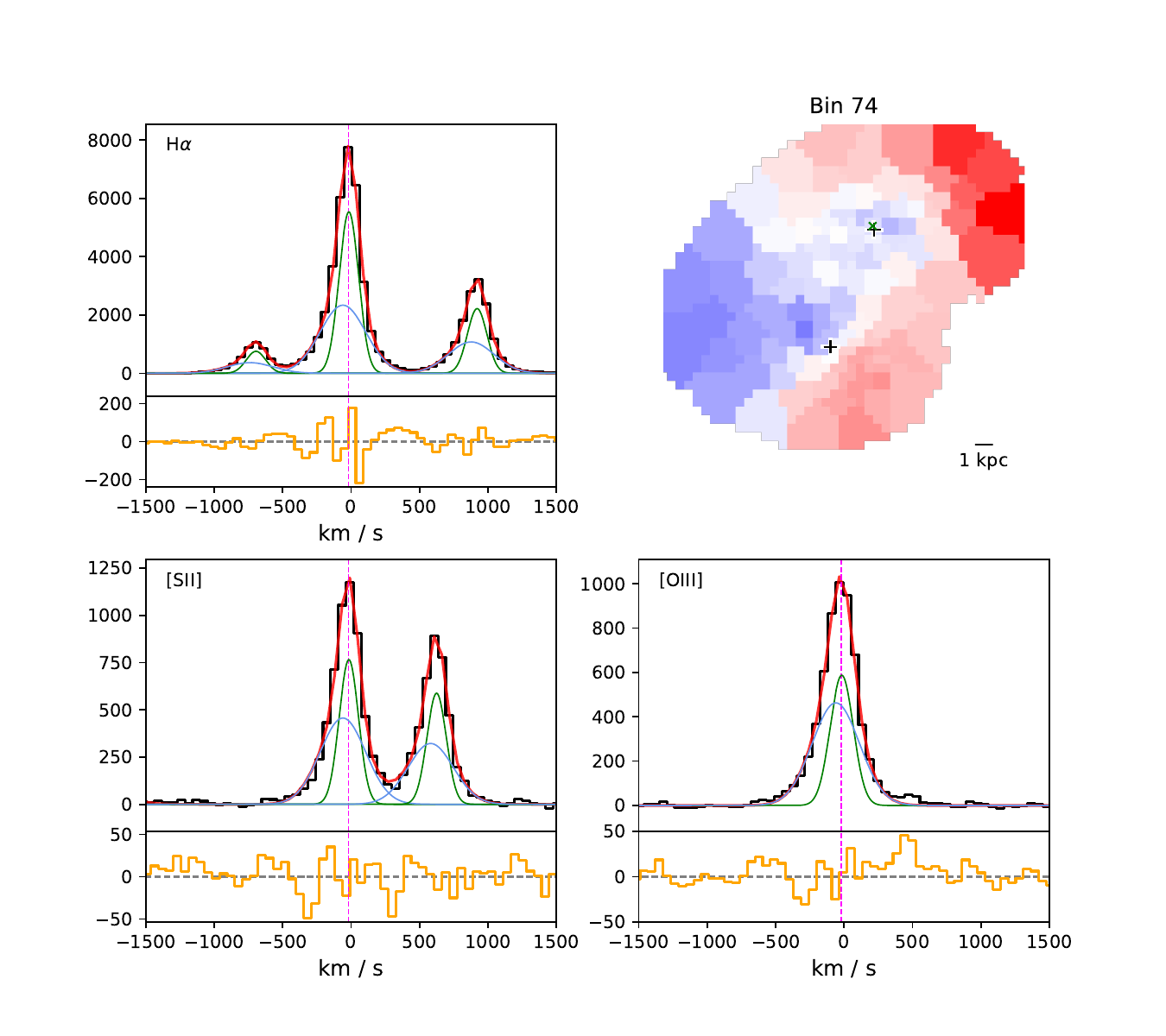}
    \caption{Examples of fits to the H$\alpha$,
      [NII]$\lambda\lambda$6548,6583, [SII]$\lambda\lambda$6716,6731,
      and [OIII]$\lambda$5007 emission lines. The continuum subtracted spectrum is shown in black, the total fit in red, the narrow line component in green, and the broad line component in blue. The magenta dashed line shows the central velocity of the stellar component in the bin. The residual of the fit is shown in orange in the small panel below the fit for each region. The position of the bin is indicated by a green x on the stellar velocity map shown in the top right of each panel. The top panel shows a bin near the SE nucleus and the bottom panel a bin near the NW nucleus.}
    \label{fig:gandalf_example_fits}
\end{figure}
To extract the stellar and ionized-gas kinematics as well as the nebular emission fluxes in the MUSE data for IRAS20100-4156 we follow a similar procedure as used in \citet{sarziFornax3DProjectOverall2018}.
This is based on the use of pPXF for extracting the stellar kinematics and on the flexibility afforded by \texttt{GandALF}
\citep{sarziSAURONProjectIntegralfield2006} to capture the complicated ionized-gas emission in IRAS20100-4156, including the use of secondary components to recover the characteristics of outflowing material and of absorption lines for an initial fit of the NaI doublet.

\begin{figure}
    \centering
    \includegraphics[width=0.48\textwidth]{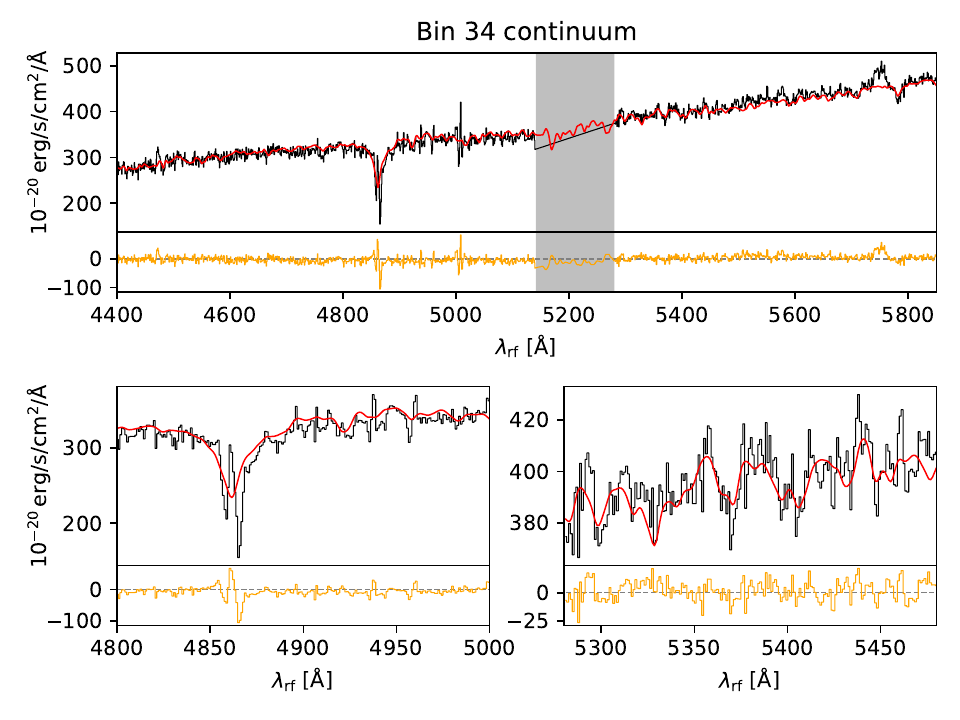}
    \includegraphics[width=0.48\textwidth]{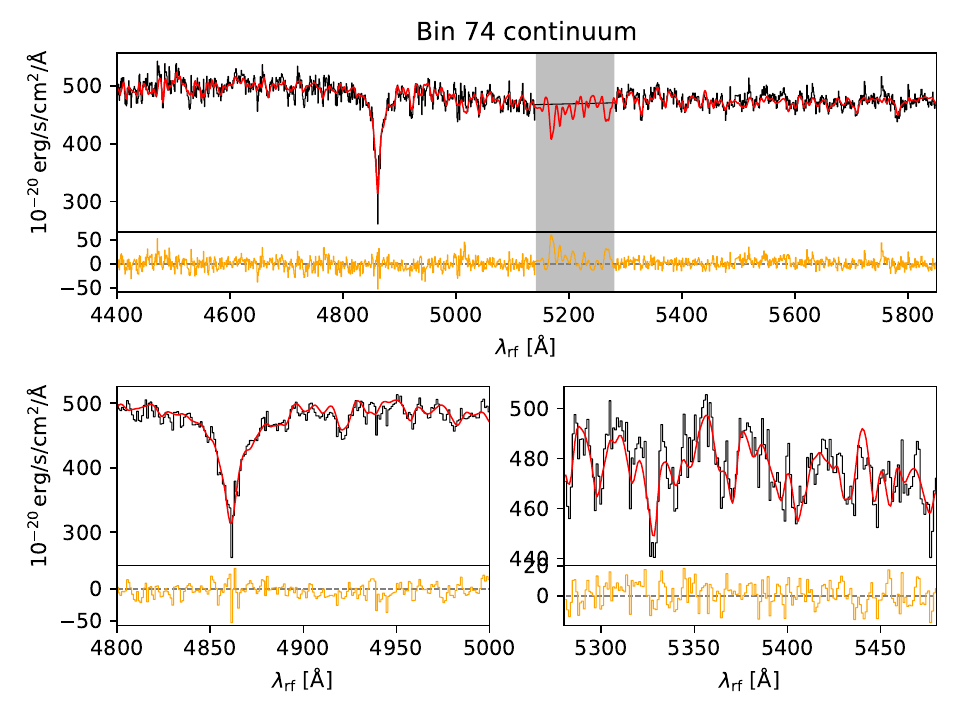}
    \caption{{\bx Examples of fits to the stellar continuum from the same bins as shown in Fig.~\ref{fig:gandalf_example_fits}. The top panel for each bin shows the full spectral range of the continuum fit, with the gray shaded range indicating the portion of the spectrum masked due to the use of adaptive optics (AO). The smaller panels on the bottom show zoom-ins of a region around the H$\beta$ absorption feature (left) and to the red side of the AO window (right) for each bin. In each panel, the spectrum with best-fit emission line templates subtracted is shown in black and the stellar template fit in red. The residual of the fit is shown in orange in the small panel below the fit for each region. The position of the bin is indicated by a green x on the stellar velocity map shown in Fig.~\ref{fig:gandalf_example_fits}. The top panel shows a bin near the SE nucleus and the bottom panel a bin near the NW nucleus.}}
    \label{fig:gandalf_cont_fits}
\end{figure}

\begin{figure}[]
    \centering
    \includegraphics[clip=true, trim=0.9cm 0.0cm 0.6cm 0.1cm,width=0.4\textwidth]{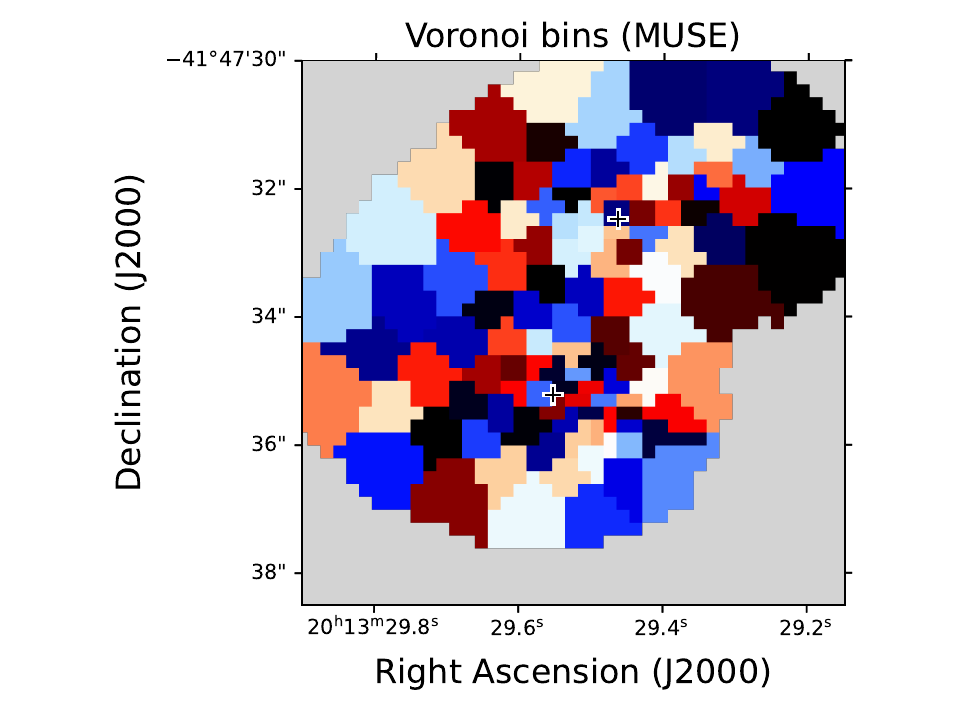}
    \caption{Map showing the Voronoi binning of the MUSE cube.}
    \label{fig:vbin_map_muse}
\end{figure}

The MUSE spectra are first fit using the IDL implementation of \texttt{pPXF} \citep{cappellariParametricRecoveryLineofSight2004} to fit the absorption-lines rich part of the spectra in the 4400\AA-5850\AA\ rest-frame wavelength range, while masking regions affected by nebular emission (as defined later by the \texttt{GandALF} fits in this spectral region) and residuals from sky lines subtractions (e.g., around 5577\AA). 
To model the stellar continuum, we used the stellar population templates from the MILES library \citep{vazdekisEvolutionaryStellarPopulation2010} and 10$^\mathrm{th}$-order additive polynomials. 
This fits returns the stellar line-of-sight velocity distribution for the given MUSE spectrum, which we restricted to only two orders giving us average velocities and velocity dispersions.
These stellar kinematics is then used when re-fitting the MUSE spectra with \texttt{GandALF}, which uses a) the same MILES templates to model the stellar continuum, b) two sets of Gaussians to model the nebular emission (respectively denoted as main and wing components from now on), c) two Gaussian absorption features to model the NaI$\lambda\lambda$5890,96 doublet absorption and d) two separated reddening components, one acting as a dust screen affecting the entire spectrum and another mimicking dust inside gas-emitting regions and thus affecting only the nebular spectrum.
The \texttt{GandALF} fit extends to a broader rest-frame fitting range between 4250\AA\ and 6850\AA\ to include all major ionized-gas emission lines from the H$\gamma$ Balmer line at 4340\AA\ to the
[SII]$\lambda\lambda$6716,6731 doublet.
All main components, used to characterize the more relaxed ionized gas, are fitted with the same kinematic parameters consisting of central velocity and line width, while their amplitudes are left to vary independently, except in cases of doublets with line ratios determined by atomic physics (such as Balmer lines and the [OIII]$\lambda\lambda$4959,5007, [OI]$\lambda\lambda$6300,6364, and [NII]$\lambda\lambda$6548,6583 doublets).
We take the same approach for the wing components used to fit outflowing material and the two absorption components used to characterize the NaI doublet, each sharing a distinct set of lines velocities and velocity dispersions.
The main and wing emission component fits are initialized, respectively, at the systemic velocity derived by the stellar kinematics and at negative 500 \kms\  velocity offset from the systemic velocity.
This initilisation is done to ensure that main and wing components do not `switch' roles during the fit, and has no significant effect on the best-fit outflow velocities in bins where an outflow is clearly detected. 
For this, we require the wing components to have an amplitude-to-noise ratio greater than three and a flux exceeding three times the line flux uncertainty of the fit.
Finally, both components in the dust model follow the \citet{calzettiDustContentOpacity2000} extinction law and E(B-V) as a fit parameter.
In practice, this \texttt{pPXF} and \texttt{GandALF} fitting procedure is used in three distinct phases in order to arrive at our final products.
In first instance, the procedure is applied to a single spectrum extracted over an aperture covering the entire galaxy (see Fig.~\ref{fig:muse_overview}). 
This is to derive the galaxy systemic velocity and optimize masks and fit initial conditions.
Second, the procedure is applied to a number of aperture spectra, extracted across different region of the IRAS20100-4156 system meant to capture variations in stellar population content (e.g., across both nuclei and most faint stellar outskirts) or dramatic changes in nebular behavior, as informed from a preliminary inspection of the MUSE cubes. 
This is done not only for further optimization but also to derive optimal stellar-population templates for each aperture, as obtained from the weighted average of the MILES model using the best-fitting returned weights. 
This effectively allows us to obtain intrinsic galactic stellar templates in the regions sampled by our apertures, which we can use to greatly speed subsequent fits across IRAS20100-4156.
This is precisely what we do in our third and last step, where we apply our procedure on individual Voronoi-binned MUSE spectra obtained after aiming at a target signal-to-noise ratio (S/N) of 40, while using such optimal stellar templates instead of the entire MILES library. 
A S/N=40 yields a good compromise of maintaining spatial resolution while obtaining sufficient signal in each bin to reliably recover fainter emission lines and components. 
Here, the S/N for each spaxel is defined as the median flux density of the spectrum in the wavelength range used in the initial \texttt{pPXF} fit divided by the median standard deviation of the spectrum over the same wavelength range.
Examples of the emission line fit for two bins near the SE and NW nuclei are shown in Fig.~\ref{fig:gandalf_example_fits}.
Figure~\ref{fig:gandalf_cont_fits} shows the continuum fit for the same example bins.
The map of the Voronoi binning is shown in Fig.~\ref{fig:vbin_map_muse}.

\section{Sodium absorption line fitting and deriving outflow properties of the neutral atomic phase}
\label{app:Na_fitting}
In the full-spectrum fit to the MUSE data, the NaI doublet is modeled with two Gaussians.
These two components are forced to share the same kinematics (central velocity and velocity dispersion) which are however independent of the kinematics used for the optical emission line templates in the fit, as they trace a different phase of the gas.
From this fit to the NaI doublet we determine the presence of blueshifted absorption, which is indicative of a outflow in the neutral atomic phase on a bin-by-bin basis.
This models is sufficient to account for the presence of nebular absorption qualitatively, but it does not allow for the accurate determination of column densities and neutral gas masses.
We therefore re-fit the region around the NaI doublet absorption feature with a more accurate model based on \citet{rupkeOutflowsInfraredLuminousStarbursts2005} and \citet{satoAEGISNATUREHOST2009}.
In this model, the optical depth due to line absorption as a function of wavelength is given by
\begin{equation}
\label{eq:tau_sodium}
    \tau (\lambda)=\tau_0 e^{-(\lambda-\lambda_0)^2/(\lambda_0 b/c)^2},
\end{equation}
where $\lambda_0$ is the central observed wavelength of the line, $\tau_0$ its central optical depth, and $b$ its Doppler parameter.
The observed intensity profile including NaI doublet absorption is then
\begin{equation}
\label{eq:intensity_sodium}
    I(\lambda) = (1 - C_f[e^{-\tau_\mathrm{blue}-\tau_\mathrm{red}}])\cdot I_0(\lambda),
\end{equation}
where $\tau_\mathrm{blue}$ and $\tau_\mathrm{red}$ are the optical depths due to absorption from the ``blue'' and ``red'' lines in the doublet respectively, $C_f$ is the covering fraction, and $I_0(\lambda)$ is the intensity profile without nebular NaI doublet absorption (i.e., stellar continuum plus HeI$\lambda$5876 emission).

\begin{figure}
    \centering
    \includegraphics[clip=true, trim=0.2cm 0.5cm 0.3cm 0.1cm,width=0.42\textwidth]{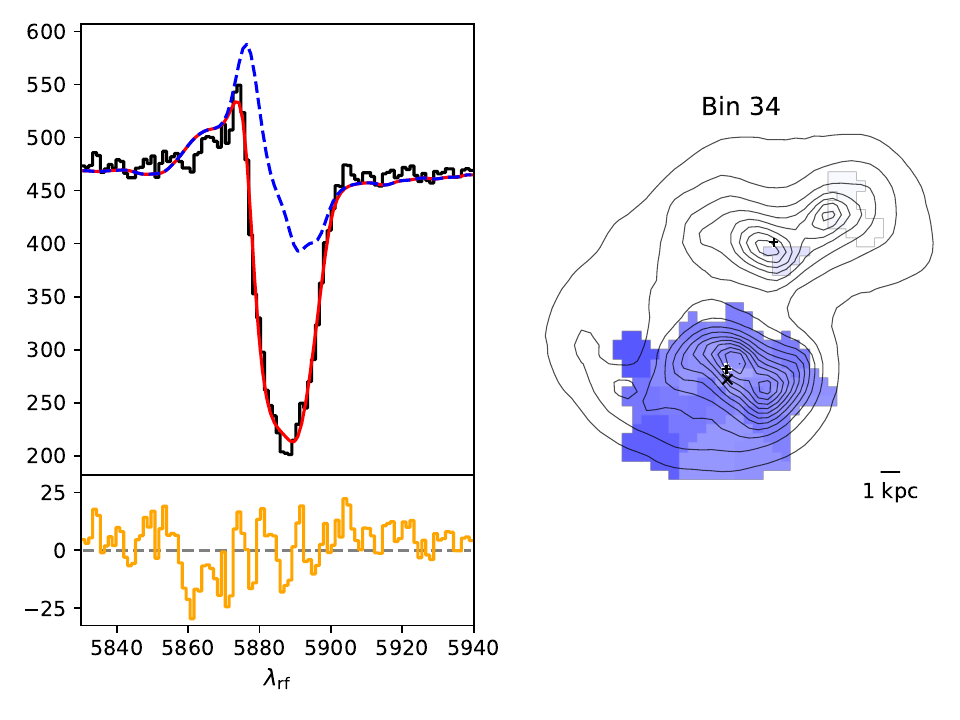}
    \includegraphics[clip=true, trim=0.2cm 0.5cm 0.3cm 0.1cm,width=0.42\textwidth]{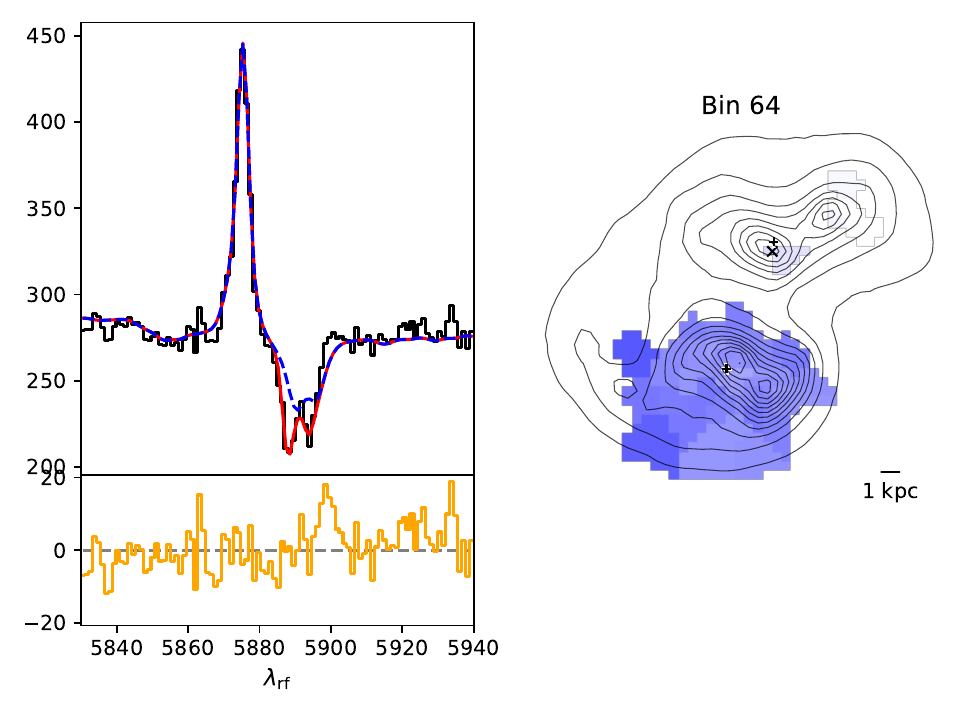}
    \caption{Examples of fits to the NaID absorption feature. The MUSE spectrum is shown in black, the total fit in red, and the best fit model without nebular NaID absorption (i.e., stars and HeI emission) in blue. The residual of the fit is shown in orange in the small panel below the fit for each region. The position of the bin is indicated by a black x on the sodium outflow velocity map shown on the right in each panel. The top panel shows a bin near the SE nucleus and the bottom panel a bin near the NW nucleus.}
    \label{fig:sodium_example_fits}
\end{figure}

We fit this model to the data using the \texttt{lmfit} python package.
To construct an intensity profile without NaI doublet absorption, we first subtract the ``placeholder'' Gaussian fit to the NaI doublet obtained with \texttt{GandALF} from the full fit.
We can then construct the intensity profile with properly modeled NaI doublet absorption using Eq.~\ref{eq:intensity_sodium}, and fit the resulting model to the MUSE data in each Voronoi bin.
The central optical depths of the two lines are linked by the ratio of their Einstein coefficients \citep{mortonAtomicDataResonance1991}, so we set $\tau_\mathrm{0,blue} = 2\tau_\mathrm{0,red}$ for this fit.
We further assume that the two doublet components share the same kinematics, so that their Doppler parameters are equal and their observed central wavelengths can be expressed via a Doppler shift from their rest frame wavelengths with a single velocity $v_\mathrm{NaI}$.
The velocity of the outflow traced by NaI absorption is then given by $v_\mathrm{NaI}-v_\mathrm{sys}$, where $v_\mathrm{sys}$ is the systemic velocity based on the redshift obtained from the \texttt{GandALF} fit to the total integrated spectrum of the galaxy.
Examples of the sodium absorption fitting are shown in Fig.~\ref{fig:sodium_example_fits}.
Using this model we can calculate the sodium column density from the best fit central optical depth and the Doppler parameter \citep{Spitzer1978book} as
\begin{equation}
    \label{eq:sodium_column}
    N(\mathrm{NaI}) = \frac{\tau_\mathrm{0,red}}{1.497\times10^{-15} f_\mathrm{red}} \left( \frac{b}{\mathrm{km/s}}\right) \left(\frac{\mathrm{\text{\AA}}}{\lambda_\mathrm{0,red}}\right)\,\mathrm{cm^{-2}}
,\end{equation}
where $\lambda_0$ and $f=0.318$ are the central wavelength in vacuum and the oscillator strength of the NaI$\lambda$5896 line.
From the sodium column density we can then calculate the column density of neutral hydrogen, following \citet{rupkeOutflowsInfraredLuminousStarbursts2005}:
\begin{equation}
    \label{eq:HI_column}
    N(\mathrm{H}) = (1-y)^{-1}10^{-(a_\mathrm{Na}+b_\mathrm{dust})}N(\mathrm{NaI}),
\end{equation}
where $y$ is the ionization fraction, $a_\mathrm{Na}$ is the galaxy's sodium abundance, and $b_\mathrm{dust}$ the depletion onto dust.
Following \citet{fluetschPropertiesMultiphaseOutflows2021}, we adopt values of $y=0.9$, $a_\mathrm{Na}=-5.69$, and $b_\mathrm{dust}=-0.95$ respectively from \citet{SavageSodiumAbundance1996} and \citet{rupkeOutflowsInfraredLuminousStarbursts2005}.
From the hydrogen column density, we can calculate the outflowing neutral atomic gas mass in each bin as
\begin{equation}
\label{eq:neutral_mass}
    M_\mathrm{HI,out} = N(\mathrm{H}) A_\mathrm{bin} C_f m_\mathrm{H},
\end{equation}
where $A_\mathrm{bin}$ is the physical area covered by the bin at the redshift of the galaxy, $C_f$ is the best-fit covering fraction, and $m_\mathrm{H}$ is the mass of the hydrogen atom.
With the mass contribution of each bin, we can calculate its contribution to the outflow mass rate analogous to the cold molecular and ionized gas, using Eq.~\ref{eq:mass_rate} with the best-fit velocity from the fit to the NaI doublet absorption.

\section{Electron density maps}
\label{app:electron_density}
\begin{figure}[!h]
    \centering
    \includegraphics[clip=true, trim=0.4cm 0.1cm 0.6cm 0.1cm,width=0.41\textwidth]{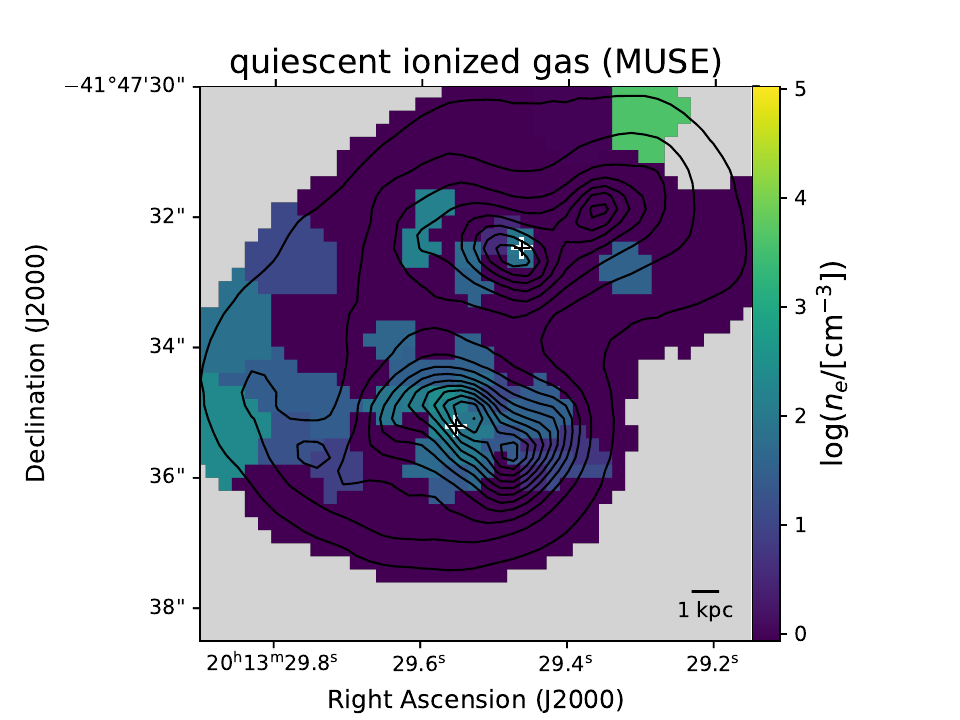}
    \includegraphics[clip=true, trim=0.4cm 0.1cm 0.6cm 0.1cm,width=0.41\textwidth]{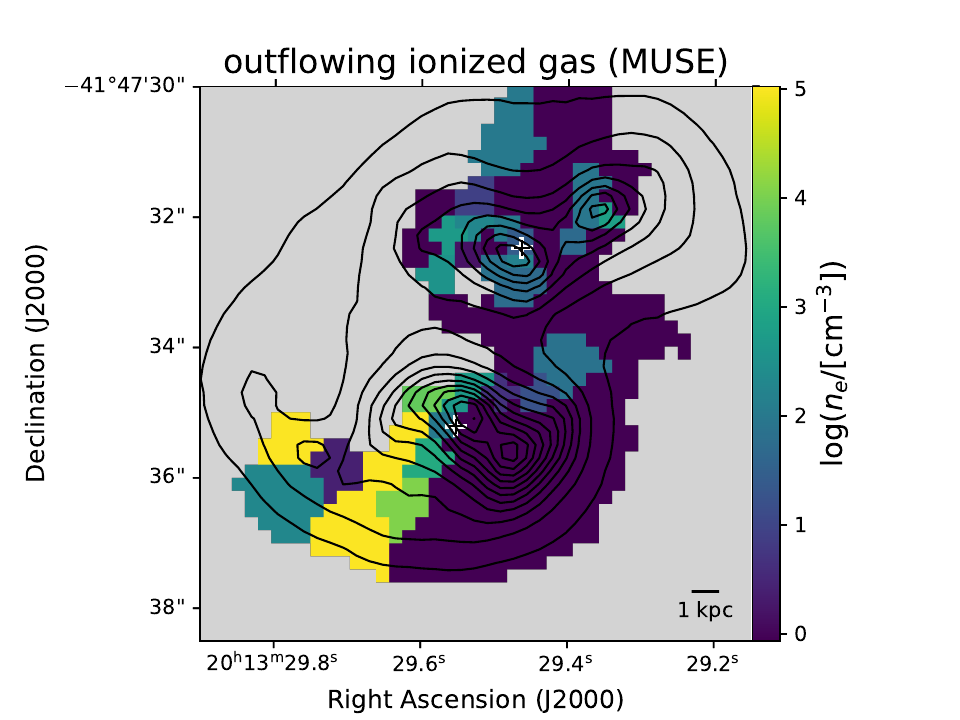}
    \caption{Map of electron density in the quiescent gas component (top) and in the outflow (bottom) for the Voronoi binned cube. Electron densities are derived from [SII] emission line ratios as detailed in Sect.~\ref{sec:results:ionized_outflow}.}
    \label{fig:electron_density_maps_vbin}
\end{figure}

\vspace{0.5cm}

\section{Gas-phase metallicity map}
\label{app:metallicity}
\begin{figure}[!h]
    \centering
    \includegraphics[clip=true, trim=0.4cm 0.1cm 0.6cm 0.1cm,width=0.41\textwidth]{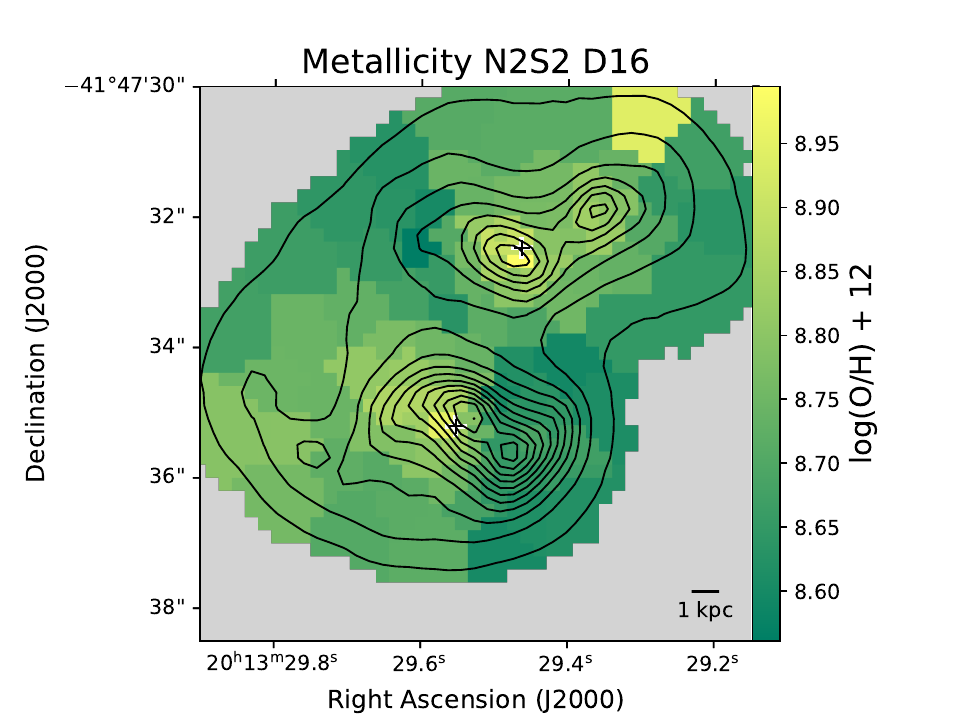}
    \caption{Map of gas-phase metallicity in the quiescent (narrow) gas component for the Voronoi binned cube. Metallicities are derived from [NII]/H$\alpha$ and [NII]/[SII] line ratio indices following \citet{dopitaChemicalAbundancesHighredshift2016}.}
    \label{fig:met_maps_vbin}
\end{figure}

\section{Correlation between dust extinction and cold gas tracers}
\label{app:dust}

\begin{figure}[]
    \centering
    \includegraphics[width=0.31\textwidth]{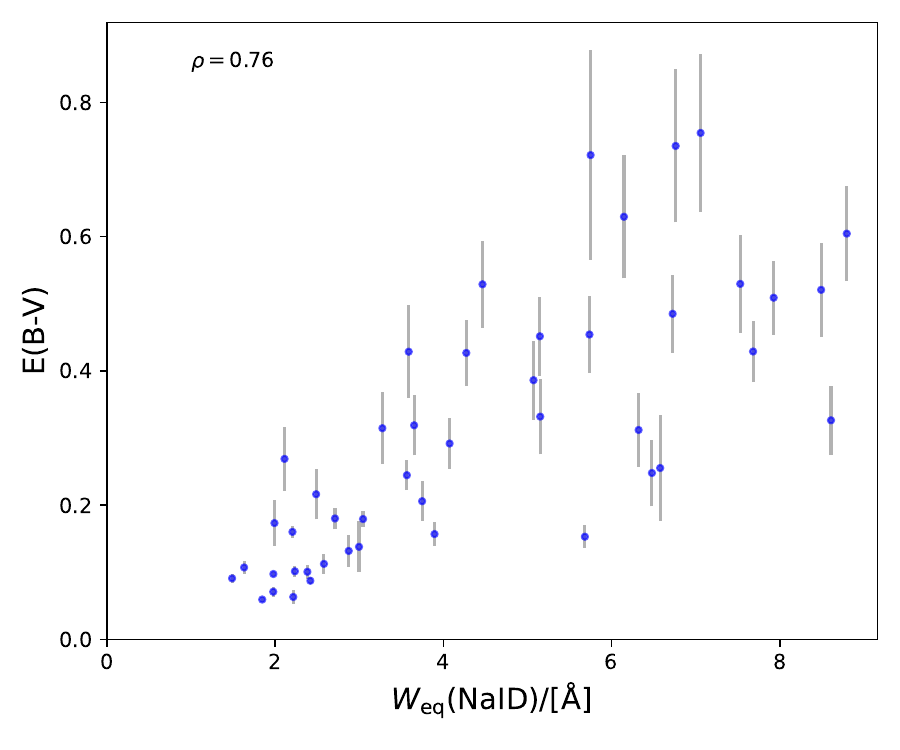}
    \includegraphics[width=0.31\textwidth]{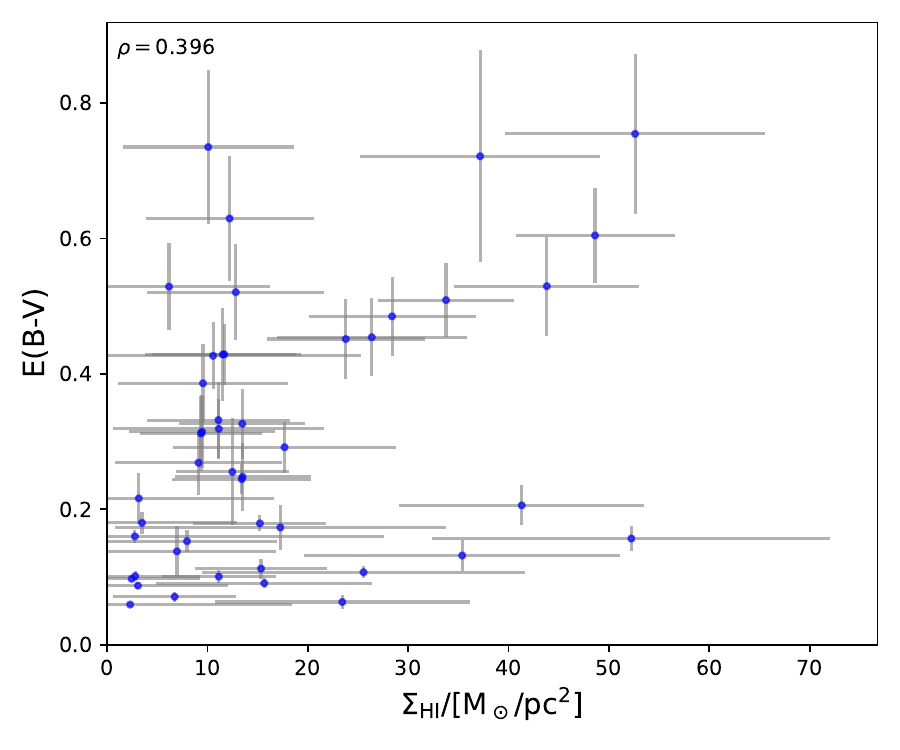}
    \caption{Measures of outflowing neutral atomic gas vs.  dust extinction E(B-V) due to the diffuse dust component used in the fit to the MUSE data. Top: Correlation with NaID absorption EW.
    Bottom: Correlation with surface density of neutral atomic hydrogen inferred from NaI absorption.}
    \label{fig:HI_dust_corr}
\end{figure}

\begin{figure}[!h]
    \centering
    \includegraphics[width=0.31\textwidth]{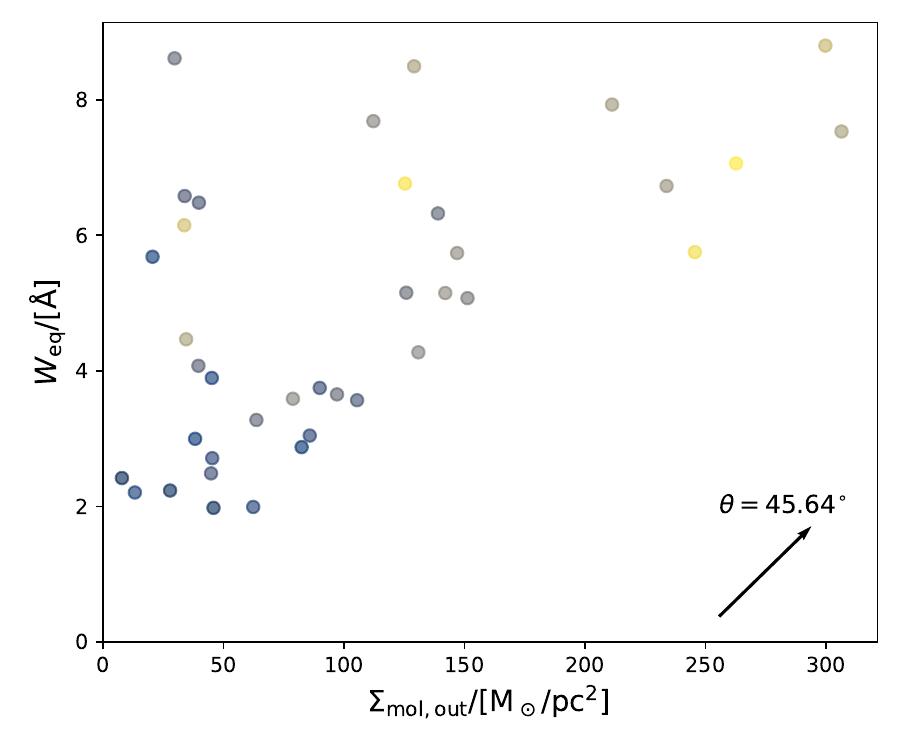}
    \includegraphics[width=0.31\textwidth]{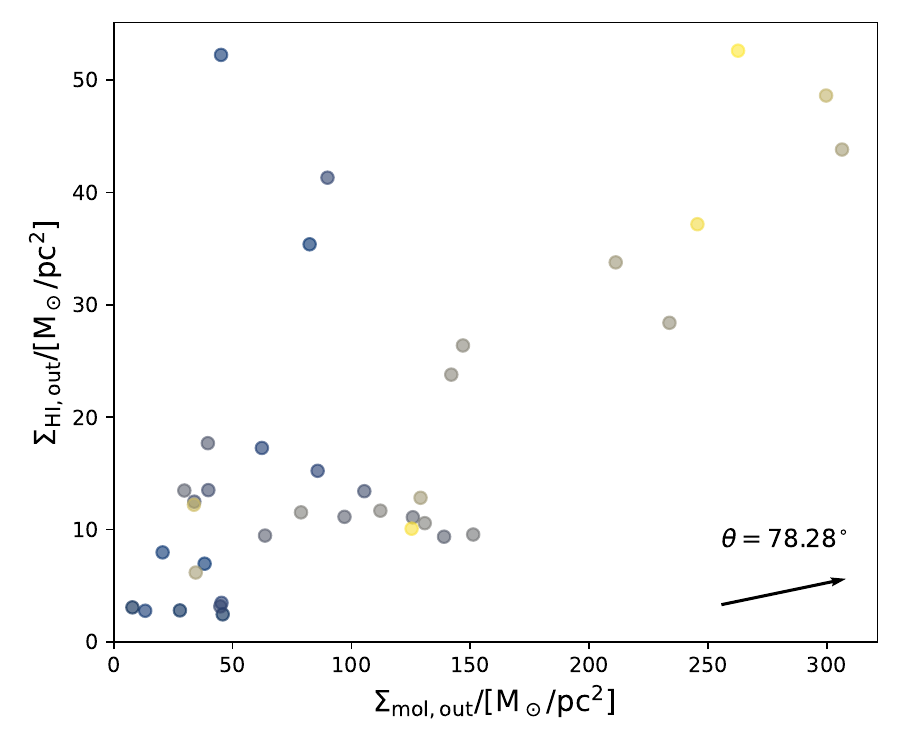}
    \caption{Molecular gas surface density in the outflow vs. NaID absorption EW (top) and neutral atomic hydrogen surface density (bottom). The color scaling corresponds to the diffuse dust extinction E(B-V) in each bin. Arrow shows the angle of maximum increase for E(B-V) in the plane.}
    \label{fig:CO_HI_dust_angle}
\end{figure}

We test our assumption that the observed diffuse dust extinction corresponds primarily to dust mixed with outflowing cold molecular gas by comparing (partial) correlation coefficients.
As is evident from comparing Figs.~\ref{fig:vorbin_neut_eqw_map} and \ref{fig:vorbin_dust_maps}, the neutral atomic phase of the outflow traced by NaID absorption and the peak in extinction E(B-V) due to diffuse dust share a similar morphology peaking to the SE of the SE nucleus.
Indeed there is a clear correlation between the EW of the NaID absorption ($W\mathrm{_{eq}(NaID)}$) and the diffuse dust extinction E(B-V), as can be seen in the top panel of Fig.~\ref{fig:HI_dust_corr}.
However, this correlation is much weaker when the measured NaID absorption is converted into a neutral hydrogen surface density ($\Sigma_\mathrm{HI,out}$), as shown in the bottom panel of Fig.~\ref{fig:HI_dust_corr}.
It is possible that this is in some part due to degeneracies between fit parameters, particularly the covering fraction and central optical depth (see Appendix~\ref{app:Na_fitting})
We use partial correlation coefficients to determine the primary scaling for diffuse dust extinction and express it as an angle of maximum increase in Fig.~\ref{fig:CO_HI_dust_angle}, adapting a method from \citet{bluckHowCentralSatellite2020}.
When comparing the surface density of outflowing cold molecular gas ($\Sigma_\mathrm{mol,out}$) and $W\mathrm{_{eq}(NaID)}$, diffuse dust extinction scales practically equivalently with both quantities with an angle of maximum increase of $45.64^{\circ}$ in the associated plane (top panel of Fig.~\ref{fig:CO_HI_dust_angle}).
This angle corresponds to the direction in which the increase in E(B-V) is steepest in the plotted parameter plane.
The same analysis using $\Sigma_\mathrm{HI,out}$ and $\Sigma_\mathrm{mol,out}$ results in an angle of $78.28^{\circ}$, showing that the cold molecular gas in the outflow is much more closely associated with diffuse dust extinction.

We use the same method to compare the outflowing cold molecular gas component to it quiescent counterpart, since diffuse dust extinction is correlated with both quiescent molecular gas surface density, ($\rho=0.666$) and outflowing molecular gas surface density ($\rho=0.744$).
This is likely due to the fact that both molecular gas components peak near the center of the galaxy, indeed the correlation between the quiescent and outflowing molecular gas surface densities is significant ($\rho=0.78$).

\begin{figure}[!h]
    \centering
    \includegraphics[width=0.42\textwidth]{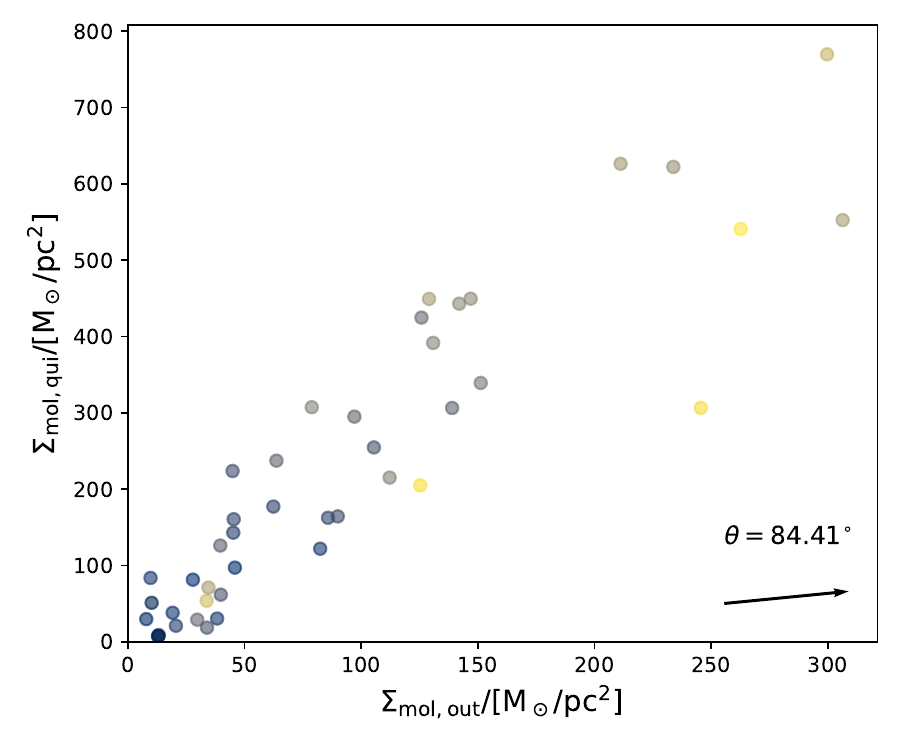}
    \caption{Molecular gas surface density in the disk vs. in the outflow. The color scaling corresponds to the diffuse dust extinction E(B-V) in each bin. The arrow shows the angle of maximum increase for E(B-V) in this plane.}
    \label{fig:CO_dust_angle}
\end{figure}

The angle of maximum increase in the $\Sigma_\mathrm{mol,qui}$ vs. $\Sigma_\mathrm{mol,out}$ plane is $84.41^{\circ}$, closely aligned with the x-axis.
We can therefore conclude that diffuse dust extinction in IRAS20100-4156 primarily arises from outflowing cold molecular gas.

\section{CO(1-0) emission line fitting examples}
\label{app:co_line_fitting}

\begin{figure}[!h]
    \centering
    \includegraphics[clip=true, trim=1.5cm 0.0cm 0.8cm 0.1cm,width=0.42\textwidth]{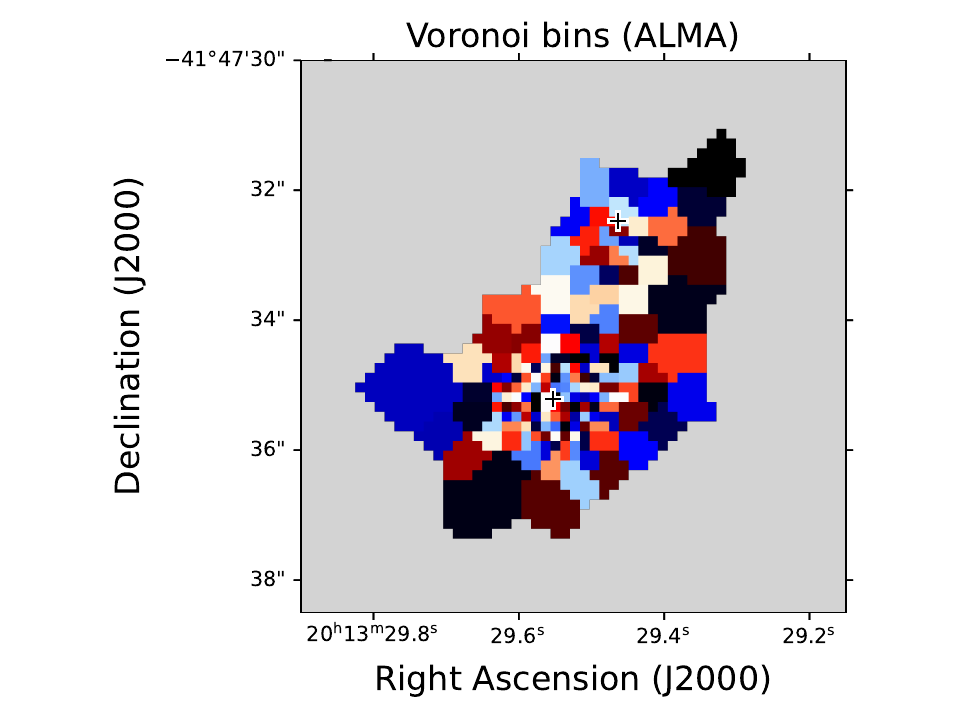}
    \caption{Map showing the Voronoi binning of the ALMA cube.}
    \label{fig:vbin_map_alma}
\end{figure}

\begin{figure*}
    \centering
    \includegraphics[clip=true, trim=0.8cm 0.1cm 1.3cm 0.1cm,width=0.48,width=0.48\textwidth]{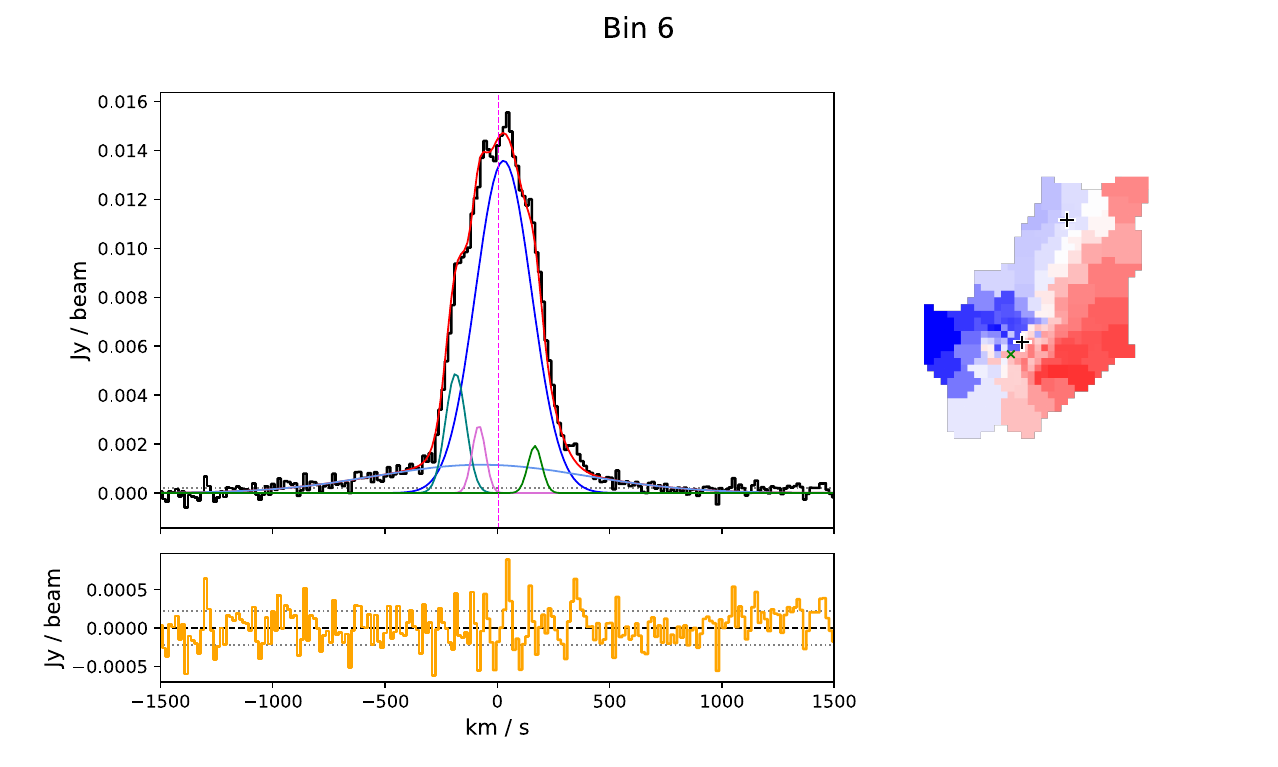}
    \includegraphics[clip=true, trim=0.8cm 0.1cm 1.3cm 0.1cm,width=0.48,width=0.48\textwidth]{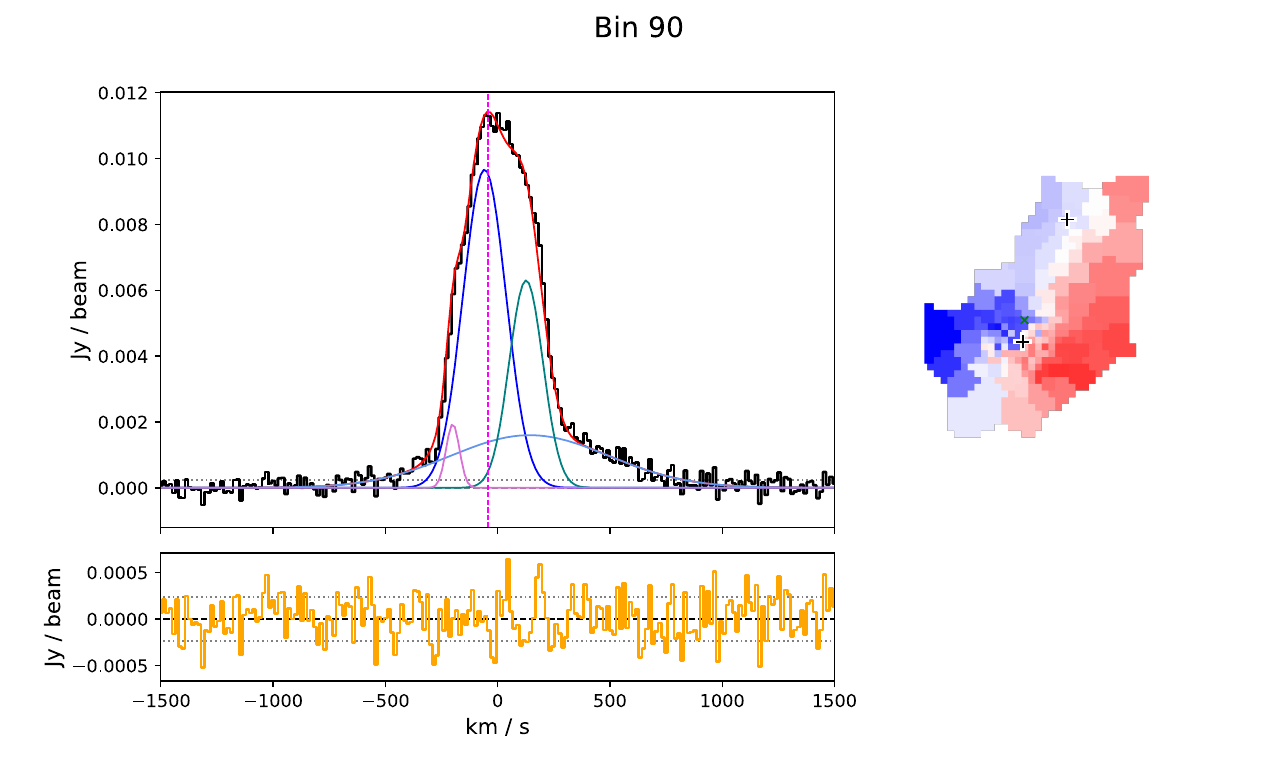}
    \includegraphics[clip=true, trim=0.8cm 0.1cm 1.3cm 0.1cm,width=0.48,width=0.48\textwidth]{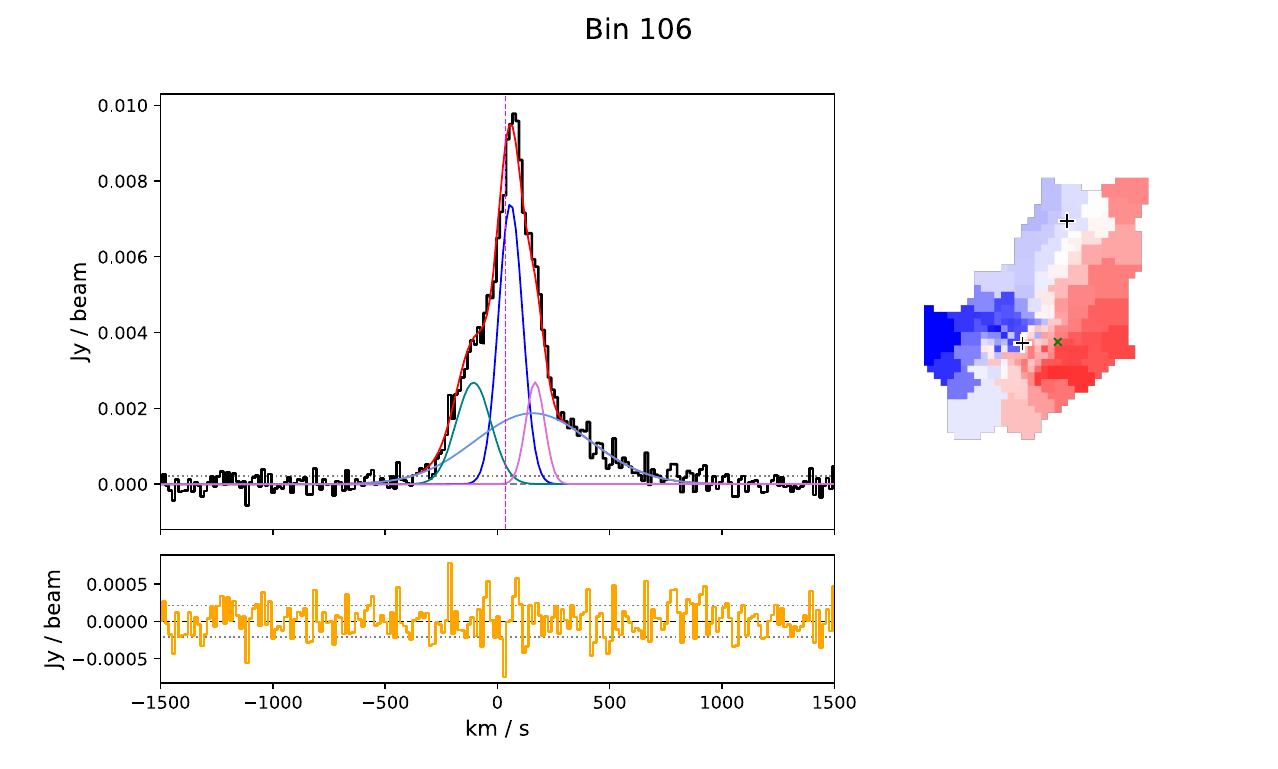}
    \includegraphics[clip=true, trim=0.8cm 0.1cm 1.3cm 0.1cm,width=0.48,width=0.48\textwidth]{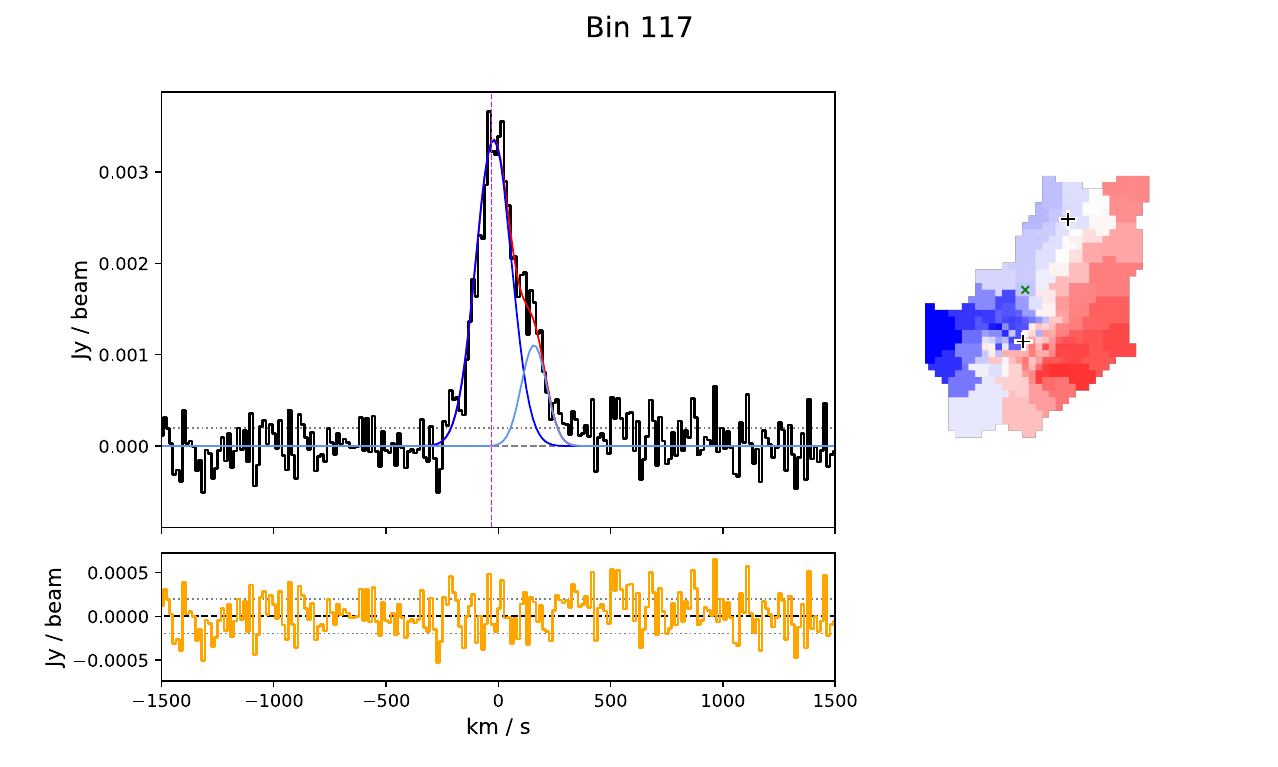}
    \includegraphics[clip=true, trim=0.8cm 0.1cm 1.3cm 0.1cm,width=0.48,width=0.48\textwidth]{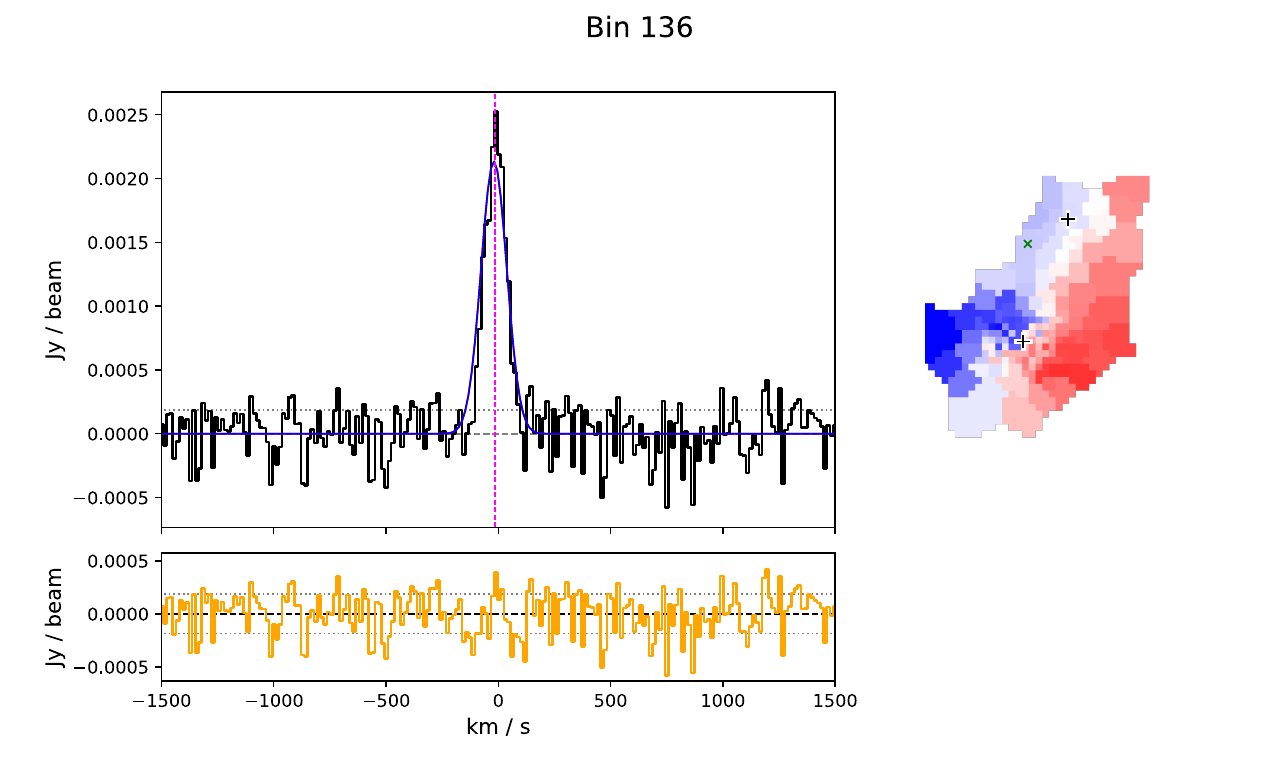}
    \includegraphics[clip=true, trim=0.8cm 0.1cm 1.3cm 0.1cm,width=0.48,width=0.48\textwidth]{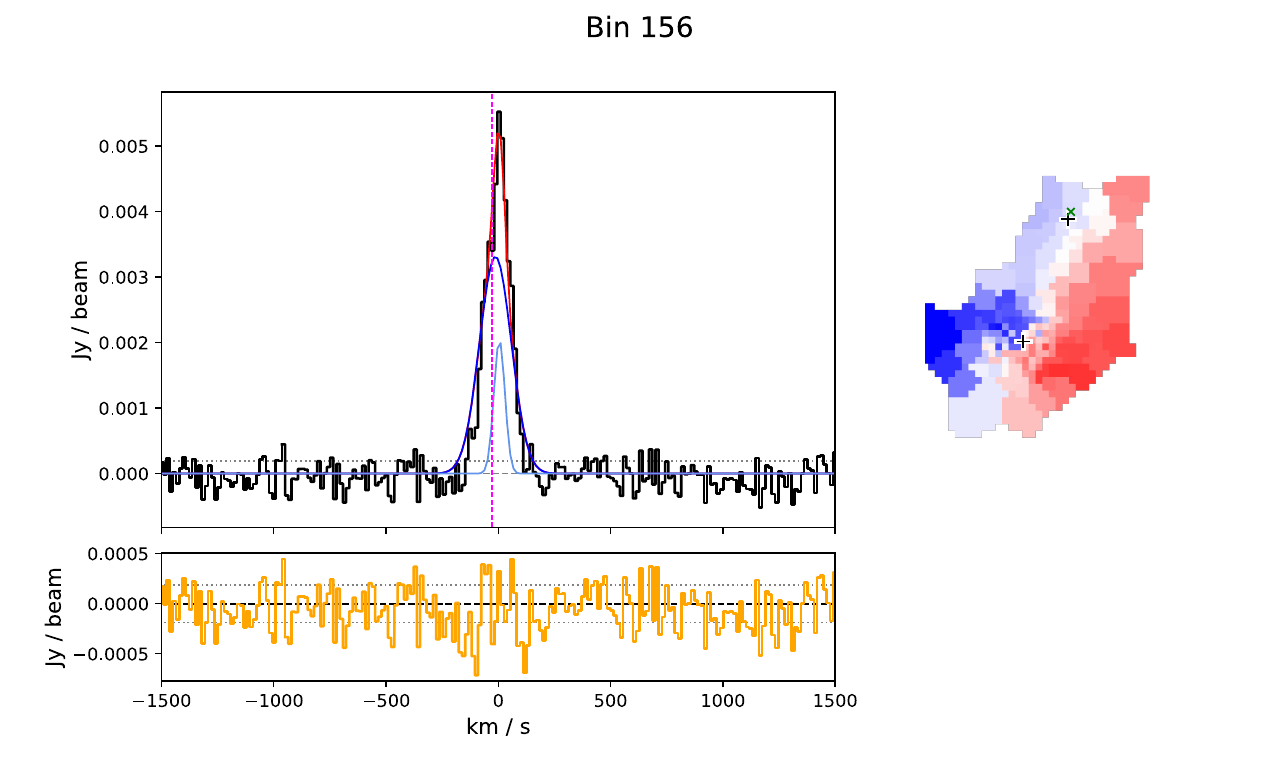}
    \caption{Examples of constrained multi-Gaussian fits used in the outflow identification process as described in Sect.~\ref{sec:methods:outflow_identification}. In black is the integrated spectrum extracted from a given Voronoi bin, in blue the gaussian component constrained to trace the quiescent gas following stellar kinematics, light blue, teal pink and green lines show additional components added to the fit in cases of complex line profiles. The magenta dashed line shows the central velocity of stars in the bin. The gray dotted line shows the noise level of the spectrum. The red solid line shows the sum of all components. The bottom panel shows the residual of the fit (data minus sum of all fit components) in orange and the noise level again as a dotted gray line. To the right of each plot is the position of the bin marked as a green x in the velocity map of the quiescent gas.}
    \label{fig:co10_fit_examples}
\end{figure*}

Fig.~\ref{fig:vbin_map_alma} shows a map of the Voronoi bins used in the analysis of CO(1-0) emission in IRAS 20100-4156.
Figure~\ref{fig:co10_fit_examples} shows examples of constrained multi-Gaussian fits to the CO(1-0) emission in several bins across the galaxy.

\end{document}